\documentclass[onecolumn,prd,nofootinbib]{revtex4}
\usepackage{epsf}
\usepackage{dcolumn}
\usepackage{bm}
\usepackage{graphicx}

\usepackage[]{natbib}
\setcitestyle{number,square,comma}

\newcommand{\be}{\begin{equation}}
\newcommand{\ee}{\end{equation}}

\def\4he{$^4$He}
\def\3he{$^3$He}
\def\7li{$^7$Li}

\def\ltsim{\raise 2pt \hbox {$<$} \kern-1.1em \lower 4pt \hbox {$\sim$}}
\def\gtsim{\raise 2pt \hbox {$>$} \kern-1.1em \lower 4pt \hbox {$\sim$}}

\begin{document}
\bibliographystyle{abbrvnat}

\title{Evidence for AGN Feedback in Galaxy Clusters and Groups}

\author{Myriam Gitti$^{1,2,3}$, Fabrizio Brighenti$^4$, Brian R. McNamara$^{2,5}$\\}
\affiliation{$^1$INAF-Osservatorio Astronomico di Bologna, Via Ranzani
  1, I-40127 Bologna, Italy} 
\affiliation{$^2$Harvard-Smithsonian Center
  for Astrophysics, 60 Garden St., Cambridge MA 02138, USA}
\affiliation{$^3$INAF-Istituto di Radioastronomia di Bologna, Via Gobetti 101
  1, I-40129 Bologna, Italy} 
\affiliation{$^4$Dipartimento di Astronomia, Universit\`a di Bologna, via Ranzani 1, I-40127 Bologna, Italy}
\affiliation{$^5$Physics\&Astronomy Dept., Waterloo Univ.,
  200 Univ. Ave. W, Waterloo, ON, N2L 2G1, Canada }

\date{{\today}\\}

\begin{abstract}
  The current generation of flagship X-ray missions, {\it Chandra} and
  {\it XMM--Newton}, has changed our understanding of the so-called
  ``cool core'' galaxy clusters and groups. Instead of the initial
  idea that the thermal gas is cooling and flowing toward the center,
  the new picture envisages a complex dynamical evolution of the
  intra-cluster medium (ICM) regulated by the radiative cooling and
  the nongravitational heating from the active galactic nucleus (AGN).
  Understanding the physics of the hot gas and its interplay with the
  relativistic plasma ejected by the AGN is key for understanding the
  growth and evolution of galaxies and their central black holes, the
  history of star formation, and the formation of large-scale
  structures.  It has thus become clear that the feedback from the
  central black hole must be taken into account in any model of galaxy
  evolution. In this paper, we draw a qualitative picture of the
  current knowledge of the effects of the AGN feedback on the ICM by
  summarizing the recent results in this field.
\end{abstract}



\maketitle


\section{~Introduction}
\label{intro.sec}

The physics of the intra-cluster medium (ICM) of clusters and groups
of galaxies is complex.  The current generation of X-ray satellites,
\textit{Chandra} and \textit{XMM-Newton}, has shown indeed that it is
regulated by yet poorly understood non-gravitational processes beyond
simple gravity, gas dynamics and radiative cooling usually considered
in the standard cold dark matter cosmological scenario
\citep{White-Rees_1978}.  In particular, an important discovery from
high-resolution X-ray observations was that the amount of thermal gas
radiatively cooling to low temperatures is much less than what is
predicted by the standard ``cooling flow'' model (see
\citep{Fabian_1994, Peterson_2001, David_2001, Peterson-Fabian_2006}
and references therein), thus radically changing our understanding of
the so-called ``cool core'' systems.  The implication is that the
central gas must experience some kind of heating due to a feedback
mechanism that prevents cool cores from establishing cooling flows at
the rates predicted by earlier, low-resolution X-ray observations.
Establishing the source of this heating, and understanding when and
how it takes place, has become a major topic of study in extragalactic
astrophysics.

Based on observational evidence and theoretical modelling, the primary
source of feedback has been identified in the outbursts and
accompanying energy injection, likely intermittent, from the active
galactic nucleus (AGN) of the dominant cD galaxies (e.g.,
\citep{McNamara-Nulsen_2007} and references therein), which host the
most massive black holes in the local Universe.  AGNs manifest as
central radio sources, which are commonly observed in cool core
clusters \citep{Burns_1990}.  Most of the cool core systems have
highly disturbed X-ray morphologies, and radio observations clearly
show that AGN jets are the cause of many of the structures revealed by
the X-ray telescopes. Such surface brightness features, including
apparent depressions or ``cavities'' in the X-ray images and sharp
density discontinuities interpreted as shocks, indicate a strong
interaction between the central AGN and the intra-cluster medium
(ICM).  The incidence and variety of cavities, shocks, and ripples
observed both in the radio and in X-rays in the hot ICM provides
direct evidence of the widespread presence of AGN-driven phenomena
(e.g., \citep{Birzan_2008,Giacintucci_2011b} for sample studies of
clusters and groups, respectively).

Such AGN feedback has a wide range of impacts, from the formation of
galaxies, through to the explanation of the observed relation between
the black hole mass and the bulge velocity dispersion (which indicates
a causal connection or feedback mechanism between the formation of
bulges and their central black holes, e.g., \citep{Magorrian_1998}),
to the regulation of cool cores which explains why cooling and star
formation still proceeds at a reduced rate.  On the other hand, the
details of how the feedback loop operates are still unknown.  Feedback
is also required to suppress the overproduction of massive galaxies
predicted by dark-matter-only simulations and to break the
self-similarity of clusters
\citep[e.g.,][]{Benson_2003,Croton_2006,Balogh_2006}.  The nature of
this feedback is therefore vital to our understanding of galaxy
evolution (\citep {Cattaneo_2009} and references therein).

After a brief discussion of the importance of galaxy clusters and
their scaling relations for the study of cosmic evolution (\S
\ref{cosmo.sec}), we overview the role of AGN feedback in structure
formation (\S \ref{role.sec}) and the basic properties of clusters in
X-rays emphasizing the hot intra-cluster medium (\S \ref{icm.sec}).
We then focus on the observational evidence of AGN feedback in action
in galaxy clusters and groups (\S \ref{feedback.sec}) and finally give
our conclusions (\S \ref{conclusion.sec}).  {\it The present paper
  does not intend to be a comprehensive review, but aims instead at
  drawing a qualitative picture of the impact of the AGN feedback on
  the ICM addressed particularly to the novices in this field.}
General reviews of clusters from an X-ray perspective were given by
\citet{Sarazin_1988} and more recently by \citet{Mushotzky_2004} and
\citet{Arnaud_2005}. A review of clusters as cosmological probes was
given by \citet{Voit_2005}, and cold fronts and shocks associated with
cluster mergers were reviewed by \citet{Markevitch-Vikhlinin_2007}.
An exhaustive review of the issues of AGN heating in the hot
atmospheres was recently given by \citet{McNamara-Nulsen_2007}.

Throughout the paper we assume a cosmology with $H_0 = 70 \mbox{ km
  s}^{-1} \mbox{ Mpc}^{-1}$ and $\Omega_M=1-\Omega_{\Lambda}=0.3$,
where not specified otherwise.


\section{~Clusters of Galaxies as cosmological probe}
\label{cosmo.sec}

The existence of clusters of galaxies and of other cosmic structures
demonstrates that the Universe is not perfectly homogeneous.  The
matter density of the primordial Universe must have been slightly
inhomogeneous, with overdense perturbations which deviate from the
mean density. In the so-called ``Concordance Model'' largely accepted
today as the standard cold dark matter cosmological scenario, cosmic
structures like galaxies and clusters of galaxies originate from the
gravitational instability of these primordial density fluctuations.
The formation of structures from perturbations in the density
distribution of cold dark matter is a hierarchical process. Small
subclumps of matter are the first to deviate from the Hubble flow, to
collapse and to experience gravitational relaxation because the
density perturbations have larger amplitudes on smaller mass scales.
These small objects then undergo a merging process to form larger and
larger structures, up to the clusters of galaxies
\citep{White-Rees_1978}.

Galaxy clusters trace the high-density tail of the primordial field of
dark matter density perturbations, and their numerical density as a
function of redshift, $z$, is highly sensitive to the specific
cosmological scenario (e.g., \citep{Rosati_2002} and references
therein).  Therefore, if one builds the so-called ``cluster mass
function'' $n_M(M)$, i.e. the number density of clusters with mass
greater than $M$ in a comoving volume element, and determines its
evolution with redshift, it is possible to constrain the main
cosmological parameters from the comparison between the observations
and the theory predictions.  A complete understanding of the details
of the process of hierarchical merging would require accurate
numerical simulations. However, many fundamental aspects can be
illustrated by spherically-symmetric, simplified models of cluster
formation. In particular, the combination of spherical top-hat
collapse models with the growth function for linear perturbations
(assumed to be gaussian) has led to a variety of semi-analytical
methods to express the cluster mass function in terms of cosmological
parameters (the seminal work in this field is by
\citep{Press-Schechter_1974}).

Therefore, the comparison between the theoretical mass function and
the mass function determined from observations allows one to constrain
the main cosmological parameters, although with a degeneracy between
the matter density parameter, $\Omega_M$, and the power spectrum
normalization of the perturbations within a comoving sphere of radius
8 $h^{-1}$ Mpc, $\sigma_8$.  Such a degeneracy can be broken by
studying the redshift evolution of the mass function, which is highly
sensitive to $\Omega_M$, by taking into account the evolution of the
observables (see the reviews by \citep{Rosati_2002, Voit_2005,
  Borgani_2006} and references therein). The accuracy of the
cosmological parameter measurements is currently limited by
uncertainties in the relations between cluster masses and the
observable properties that trace these masses, such as luminosity or
temperature. In order to measure the mass function from a large sample
of clusters is indeed necessary to link the mass to these quantities
which are easily observable.

In this context are very useful the so-called {\it ``self-similar
  scaling relations''}, derived naturally by considering that the
cosmological structures originate from scale-free density
perturbations and that the thermodynamical properties of the ICM are
determined by scale-free gravity only \citep{Kaiser_1986}.  Under
these assumptions, galaxy clusters of different masses may be
considered as a scaled version of each other.  The density of each
dark matter halo, $\rho_{DM}$, is proportional to the critical density
of the Universe at the cluster's redshift, $\rho_{c,z}$, through the
so-called {\it ``overdensity''} $\Delta = \rho_{DM} / \rho_{c,z}$,
where $\rho_{c,z} = 3 H_z^2 / 8 \pi G$, and the expression for the
Hubble constant at redshift $z$ in a flat $\Lambda$CDM Universe is
\citep[e.g.,][]{Bryan-Norman_1998}: $H_z = H_0 \sqrt{\Omega_m (1+z)^3
  + \Omega_{\Lambda}} \equiv H_0 \; E(z)$.  Therefore, all clusters
should have the same properties when rescaled by $\Delta$.

If we define the mass $M$ as the mass $M_{\Delta}$ inside the radius
$R_{\Delta}$ at a given overdensity $\Delta$, we obtain: $ M_{\Delta}
\propto \rho_{c,z} \cdot \Delta \cdot R_{\Delta}^3 \propto \rho_{c,0}
\cdot E(z)^2 \cdot \Delta \cdot R_{\Delta}^3 $, and thus we get to the
$M-R$ relation in the form
\begin{equation}
\label{mr.eq}
R \propto M^{1/3} \cdot E(z)^{-2/3}
\end{equation}  

During cluster formation, the gravitational collapse of the diffuse
gas in the potential well of the dark matter halo heats the gas itself
at the virial temperature of the potential well that confines it: $
T_{\rm vir} \sim G M \mu m_{\rm p} / k R \sim 10^8 K $ where $M$ is
the total mass, $k$ is the Boltzmann constant, $\mu \sim 0.6$ is the
mean molecular weight and $R$ is the virial radius.  The gas is thus
heated to X-ray emitting temperatures and becomes a plasma in
hydrostatic equilibrium whose emissivity is proportional to the square
of its density (see \S \ref{icm.sec}).  The virial temperature of an
isothermal sphere of mass $M$ is: $ kT \propto M/R \propto M^{2/3}
\cdot E(z)^{2/3}$, leading to the $M-T$ relation in the form
\begin{equation}
\label{mt.eq}
M \propto T^{3/2} \cdot E(z)^{-1}
\end{equation}

From these relations it is possible to derive the relation between
temperature and luminosity emitted by the hot gas through
bremsstrahlung emission: $L_{\rm X} \propto \rho^2 \cdot \Lambda \cdot
V$, where $\rho$ is the average gas density and $\Lambda$ is the
cooling function, that in the bremsstrahlung regime is $\propto
T^{1/2}$ (see \S \ref{icm.sec}).  Assuming that the gas distribution
traces the dark matter distribution, $\rho \propto \rho_{DM} \propto
\rho_{c,z}$, we obtain: $ L_{\rm X} \propto \rho \cdot T^{1/2} \cdot M
\propto \rho_o \cdot E(z)^2 \cdot T^{1/2} \cdot M \propto E(z)^2 \cdot
T^{1/2} \cdot T^{3/2} \cdot E(z)^{-1} $ thus deriving the $L-T$
relation in the form:
\begin{equation}
\label{lt.eq}
L_{\rm X} \propto T^2 \cdot E(z) 
\end{equation}

By combining the $M-T$ relation (Eq. \ref{mt.eq}) with the $L-T$
relation (Eq.  \ref{lt.eq}), we can finally derive the $M-L$ relation
that links the mass directly to the observable luminosity: $ L_{\rm X}
\propto [M^{2/3} E(z)^{2/3}]^2 E(z)$, finding
\begin{equation}
M \propto L_X^{3/4} \cdot E(z)^{-7/4}
\end{equation}

In principle, once calibrated with simulations and/or observations,
these scaling relations provide a method to link the mass of clusters
to observables under the assumption that the process of structure
formation is guided by gravity alone.  On the other hand,
deviations from these relations testify the presence of physical
processes more complex than gravitational dynamics only, which modify
the thermodynamical properties of the diffuse baryons and therefore
the relations between observables and cluster masses.  In particular,
a number of observational measurements seems to indicate that the
$L-T$ relation is steeper that predicted by self-similar models, and
is in the form $L \propto T^{2.5 - 3}$ (e.g., \citep{Markevitch_1998,
  Arnaud-Evrard_1999, Ettori_2004, Maughan_2006, Ota_2006,
  Mantz_2010}, see also \S \ref{giant.sec}).  This observed breaking
of the scaling relation has been ascribed to the presence of some
excess entropy in the gas due to primordial nongravitational heating
before the cluster virialization \citep{Kaiser_1991,
  Evrard-Henry_1991}, and is one of the strongest evidence for
nongravitational processes acting in the ICM.

The main source of uncertainty in the determination of cosmological
parameters from studies of cluster samples arises then from the
uncertainty in the normalization, shape and evolution of the
relationships that relate the cluster masses to the observables.  In
order to understand better such relations it is essential to
investigate how the structure formation and AGN feedback affect the
evolution of what we can observe, i.e. the baryons in clusters.


\section{~Role of AGN feedback in galaxy evolution}
\label{role.sec}

One of the main problems of the current cosmological model is why so
few baryons have formed stars \citep{Cole_1991,White-Frenk_1991}.
Numerical simulations of cosmological structure formation that include
the hydrodynamics of baryons and the radiative cooling processes
predict that \gtsim 20\% of the baryons should have condensed into
galaxies, but only \ltsim 10\% have been observed in the form of stars
\citep[e.g.,][]{Balogh_2001}.  In particular, simulations that include
only gravitational heating predict an excessive cooling of baryons
that results in a population of galaxies which are too massive and too
bright with respect to the ones observed, thus failing to reproduce
the truncation of the high-luminosity end of the galaxy luminosity
function \citep{Benson_2003,Sijacki-Springel_2006}.

Instead of residing in the cD galaxies as predicted by simulations,
most baryons are observed in the hot ICM.  This problem may find a
solution in the nongravitational heating supplied by supernovae (SN)
and active galactic nuclei (AGN).  Supernovae are essential in the
process of enrichment of the ICM to the metallicity level observed
\citep{Metzler-Evrard_1994,Borgani_2002}, and from the heavy-element
abundances in clusters it is estimated that during a cluster history
they supply a total amount of energy of the order of 0.3-1
keV/particle \citep[e.g.,][]{Finoguenov_2001, Pipino_2002}.  This
  is not enough energy to quench cooling in massive galaxies
\citep{Borgani_2002}, as the energy input required to explain the
mass-observable relations is $\sim 1-2$ keV/particle
\citep{Tornatore_2003, Voit_2002, Wu_2001}.  Energetically, AGN
heating appears to be the most likely mechanism to severely reduce the
supply of gas from the hot ICM in massive galaxies and to explain the
observed entropy profiles \citep{Benson_2003, Scannapieco-Oh_2004,
  Donahue_2006, Voit-Donahue_2005, Voit_2005}.  AGNs are powered by
accretion of material onto a black hole (BH), which is located at the
center of each stellar spheroid (both bulges within spirals and
ellipticals).  Matter falling onto a black hole releases an energy of
the order of $E_{\rm BH} = \epsilon M c^2$, where $\epsilon \approx
0.1$ is the efficiency.  For supermassive black holes (SMBH) of masses
$\sim 10^9 M_{\odot}$, the amount of energy released during their
formation and growth is of the order of $ E_{\rm BH} \sim 2 \times
10^{62}$ erg s$^{-1}$.  Even a tiny fraction (\ltsim 1\%) of the
energy released within the bulge could heat and blow away its entire
gas content in small systems and prevent cooling, thus explaining the
lack of star formation in bulges.  An extraordinary discovery obtained
recently in astrophysics is the correlation between the mass of the
central black hole $(M_{\rm BH})$ and the velocity dispersion
$(\sigma)$ of the galaxy's bulge, used to estimate the mass of the
bulge itself \citep{Magorrian_1998}.
This ``Magorrian relation'' $M_{\rm BH} - \sigma$ suggests that the
large-scale properties of the galaxy and the small-scale properties of
the black hole are related. In particular, each massive galaxy seems
to host a central black hole, whose mass is $\sim 0.1-0.2$\% of the
bulge stellar mass \citep{Ferrarese-Merritt_2000, Gebhardt_2000,
  Tremaine_2002}.  Such a correlation may arise from the fact
that the central black hole is able to regulate the amount of gas
available for star formation in the galaxy. The formation of black
holes and the formation of bulges are closely linked. Therefore
supermassive black holes can have a profound influence on the
formation and evolution of galaxies. The physical process regulating
these phenomena has been called {\it ``feedback''}, and the
understanding of how it acts in detail is one of the main open issues
in extragalactic astrophysics.

Clusters of galaxies are the only locations in the Universe where we
can find an almost complete census of the intergalactic baryons and a
very good description of their thermodynamical status and of their
enrichment in heavy elements. Therefore, X-ray observations of the ICM
can provide us with new important insights into the processes of
cooling and feedback which regulate galaxy formation.


\section{Clusters of Galaxies in X-rays and Thermal ICM}
\label{icm.sec}

Clusters of galaxies are the largest virialized structures in the
Universe, with typical sizes of $\sim 2-4$ Mpc and total gravitational
masses of $\sim 10^{14}-10^{15} M_{\odot}$. They are luminous X--ray
sources, with typical luminosities ranging from a few$\times 10^{43} -
10^{46}$ erg s$^{-1}$.  As first suggested by \citet{Felten_1966}, the
X--rays from clusters are primarily thermal bremsstrahlung emission
from the diffuse ICM which fills the deep potential wells and is
heated to temperatures of $\sim 10^8$ K (where $kT = 1$ keV for
$T=1.16 \times 10^7$ K) during the process of cluster formation.


\subsection{Physical Properties of Hot Diffuse Plasma}
\label{hot.sec}

The simple assumptions which are generally made in the study of the
ionization state and X--ray line and continuum emission from a low
density, hot plasma are briefly reviewed below \citep{Sarazin_1988}.

1) The time scale for elastic Coulomb collisions between particles in
the plasma is much shorter than the age or cooling time of the plasma,
therefore the free particles are assumed to have a Maxwell--Boltzmann
distribution at the temperature $T$.  This follows from considerations
on the mean free paths of particles in a plasma without a magnetic
field.  The mean free path $\lambda_{\rm e}$ for an electron to suffer
an energy exchange with another electron via Coulomb collisions is
given by \citep{Spitzer_1956}:

\begin{equation}
\label{1.libero.cammino.eq}
\lambda_{\rm e}=\frac{3^{3/2} (k T_{\rm e})^2}{4 \pi^{1/2} n_{\rm e} {\rm e}^4 
\ln{\Lambda}}
\end{equation}
~\\
where $T_{\rm e}$ is the electron temperature, $n_{\rm e}$ is the
electron number density and $\Lambda$ is the ratio of largest to
smallest impact parameters for the collisions ($\ln{\Lambda} \approx
38$).  Eq. \ref{1.libero.cammino.eq} assumes that the electrons have a
Maxwellian velocity distribution at the electron temperature $T_{\rm
  e}$. However, it can be demonstrated that if a homogeneous plasma is
created in a state in which the particle distribution is
non--Maxwellian, elastic collisions will cause it to relax to a
Maxwellian distribution on a time scale determined by the mean free
paths \citep{Spitzer_1956, Spitzer_1978}. Electrons will achieve this
equilibrium on a time scale given roughly by $t_{\rm eq}(e,e) \equiv
\lambda_{\rm e} / \langle v_{\rm e} \rangle_{\rm rms}$, where $\langle
v_{\rm e} \rangle_{\rm rms}$ is the rms electron velocity $=(3 k
T_{\rm e}/m_{\rm e})^{1/2}$:

\begin{equation}
t_{\rm eq}(e,e) \approx 3.3 \times 10^5 \mbox{ yr} 
\left(\frac{T_{\rm e}}{10^8 \mbox{ K}}\right)^{3/2} 
\left(\frac{n_{\rm e}}{10^{-3} \mbox{ cm}^{-3}}\right)^{-1}
\end{equation}
~\\
The time scale for Coulomb collisions between protons to bring them
into kinetic equilibrium is about $t_{\rm eq}(p,p) \approx (m_{\rm
  p}/m_{\rm e})^{1/2} t_{\rm eq}(e,e)$, roughly 43 times longer than
that for electrons.  After this time, the electrons and ions
(generally assumed to be protons) would each have Maxwellian
distribution, but generally at different temperatures, respectively
$T_{\rm e}$ and $T_{\rm i}$.  The time scale for the electrons and
ions to reach equipartition $T_{\rm e} = T_{\rm i}$ is $t_{\rm
  eq}(p,e) \approx (m_{\rm p}/m_{\rm e}) t_{\rm eq}(e,e)$, and for
typical values of the ICM temperature and density is $t_{\rm eq}(p,e)$
\ltsim $6 \times 10^8$ yr.  Since this is shorter than the age of the
clusters or their large-scale cooling time (although it is comparable
to or longer than the cooling time in the cores of clusters, see
Eq. \ref{tcool.eq} below), the intra-cluster plasma can generally be
characterized by a single kinetic temperature $T = T_{\rm e} = T_{\rm
  i}$, which determines the rates of all excitation and ionization
processes.  It is important to note that the mean free paths, i.e.:

\begin{equation}
\lambda_{\rm e} = \lambda_{\rm i} \approx 23 \mbox{ kpc} 
\left(\frac{T_{\rm e}}{10^8 \mbox{ K}}\right)^2 
\left(\frac{n_{\rm e}}{10^{-3} \mbox{ cm}^{-3}}\right)^{-1}
\end{equation}
~\\
are generally much shorter than the length scales of interest in
clusters ($\approx$ 1 Mpc), and therefore the ICM can be treated as a
collisional fluid, satisfying the hydrodynamic equations.

2) At these low densities, collisional excitation and de--excitation
processes are much slower than radiative decays, therefore any
ionization or excitation process is assumed to be initiated from the
ground state of an ion.  Three (or more) body collisional processes
are ignored because of the low density.

3) Stimulated radiative transitions are not important, since the
radiation field in the ICM is sufficiently dilute.

4) At these low densities, the gas is optically thin and the transport
of the radiation field can therefore be ignored.

Under these conditions, ionization and emission result primarily from
collisions of ions with electrons, and collisions with other ions can
be ignored.  The time scales for ionization and recombination are
generally considerably shorter than the age of the cluster or any
relevant hydrodynamic time scale, therefore the plasma is assumed to
be in ionization equilibrium \citep[e.g.,][]{Smith-Hughes_2010}.  The
equilibrium ionization state of a diffuse plasma depends only on the
electron temperature: since in nearly all astrophysical plasmas most
of the electrons originate in hydrogen and helium atoms, and these are
fully ionized under the conditions considered here, the ICM is
generally treated as a fully ionized plasma.


By indicating with $X$, $Y$, $Z$ the mass fraction of hydrogen,
helium, and heavier elements, respectively, the corresponding atom
number densities can be written in the form:
$n_{\rm H} = n_{\rm p} \equiv \rho X / m_{\rm p}$,
$n_{\rm He} = \rho Y / 4 m_{\rm p} = n_{\rm p} Y / 4 X$,
$n_{\rm z} = \rho Z / A m_{\rm p} = n_{\rm p} Z / A X$,
where $\rho$ is the gas density, $m_{\rm p}$ the proton mass and $A$
the mean atomic mass number (i.e., the number of nucleons) of heavier
elements.  Assuming that the gas pressure $p= n k T$ is contributed
only by electrons and protons, thus neglecting nuclei ($n = n_{\rm e}
+ n_{\rm p}$), it is possible to derive the electron density $n_{\rm
  e}$ in terms of the proton density $n_{\rm p}$. From the expression
of the number of particles contributing to the pressure, $n = 2 n_{\rm
  H} + 2 n_{\rm He} + \frac{1}{2} A n_{\rm z}$, one obtains:

\begin{equation}
n = \left(2+ \frac{1}{2} \frac{Y}{X} + 
\frac{1}{2} \frac{Z}{X} \right) n_{\rm p}
\end{equation} 
~\\
which for solar abundances ($X=0.71$, $Y=0.265$, $Z=0.025$) leads to
$n_{\rm e} \sim 1.2 n_{\rm p}$.  It is also possible to calculate the
mean molecular weight in {\it amu}, $\mu$, such that the total number
density of particles (electrons, protons and ions) is $\overline{n} =
\rho/ \mu m_{\rm p}$.  From the expression $\overline{n}= 2 n_{\rm H}
+ 3 n_{\rm He} + (\frac{1}{2} A + 1) n_{\rm z}$, in the approximation
$A \gg 1$, one obtains:

\begin{equation}
\mu = \left(2 X+ \frac{3}{4} Y + \frac{1}{2} Z \right)^{-1}
\end{equation}
~\\
which for solar abundances leads to $\mu \sim 0.6$.


\subsection{X--ray Emission from the ICM}
\label{1.xray.sec}

The X--ray continuum emission from a hot diffuse plasma, such as the
ICM, is due primarily to two processes: thermal bremsstrahlung
(free--free emission) and recombination (free--bound) emission.
Processes that contribute to X--ray line emission (bound--bound
radiation) from a diffuse plasma include collisional excitation of
valence or inner shell electrons, radiative and dielectric
recombination, inner shell collisional ionization and radiative
cascades following any of these processes.

At the high temperatures typical of clusters (in particular at $kT$
\gtsim 2.5 keV), thermal bremsstrahlung is the predominant X--ray
emission process.  The bremsstrahlung emissivity at a frequency $\nu$
(defined as the emitted energy per unit time, frequency and volume) of
a plasma with temperature $T$, electron density $n_e$ and ion density
$n_i$ is given by \citep[e.g.,][]{Rybicki-Lightman_1979}:

\begin{equation}
\label{1.em.brem.eq}
J_{\rm br}(\nu, T) = 6.8 \times 10^{-38} Z^2 n_{\rm e} n_{\rm i} 
T^{-1/2} {\rm e}^{-h \nu/k T} \overline{g}(\nu,T)
\end{equation}
~\\
where the Gaunt factor $\overline{g}(T)$, which corrects for quantum
mechanical effects and for the effect of distant collisions, is a
slowly varying function of the parameters \citep{Karzas-Latter_1961,
  Kellogg_1975}.  If the ICM is mainly at a single temperature, then
Eq.  \ref{1.em.brem.eq} indicates that the X--ray spectrum should be
close to an exponential of the frequency, as is generally observed.

The total power per unit volume emitted by thermal bremsstrahlung is
obtained by integrating Eq. \ref{1.em.brem.eq} over frequency,
obtaining:

\begin{equation}
\label{1.brem.eq}
J_{br}(T) = 1.4 \times 10^{-27} n_{\rm e} n_{\rm i} T^{1/2} Z^2 \overline{g}(T)
\end{equation}
~\\
where $\overline{g}(T)$ is a frequency average of
$\overline{g}(\nu,T)$ and is in the range 1.1 to 1.5 (choosing a value
of 1.2 will give an accuracy in the estimate of $J_{br}(T)$ to within
about 20\%, \citep{Rybicki-Lightman_1979}).  For solar abundances, the
emission is primarily from hydrogen and helium.

Compilations of the different emissivities for X--ray lines and
continua can be found in the literature
\citep[e.g.,][]{Raymond-Smith_1977, Mewe-Gronenschild_1981}.  Early
detailed calculations of the X--ray spectra predicted by different
models of the ICM have been given by \citet{Sarazin-Bahcall_1977} and
\citet{Bahcall-Sarazin_1977, Bahcall-Sarazin_1978}. In these models
most of the X--ray emission is thermal bremsstrahlung continuum, and
the strongest lines (highest equivalent width) are in the 7 keV iron
line complex.

The frequency--integrated total emissivity at a temperature $T$ can be
written as:
\begin{equation}
\label{1.em.x.eq}
J_{\rm X}(T)= \Lambda(T) n_{\rm e} n_{\rm p}  \; \;    
\mbox{ erg s}^{-1} \mbox{ cm}^{-3}
\end{equation}
where $\Lambda(T)$ is the \textit{cooling function}, which depends
upon the mechanism of the emission and can be represented as:
\begin{equation}
\label{1.cool.function.eq}
\Lambda(T)= l \, T^{\alpha}
\end{equation}
where $-0.6\,$ \ltsim $\, \alpha$ \ltsim $0.55$; for thermal
bremsstrahlung it is $l \sim 2.5 \times 10^{-27}$ and $\alpha = 1/2$
\citep[][]{McKee-Cowie_1977}.  The general behaviour of the cooling
function was calculated and discussed by
\citet{Sutherland-Dopita_1993}.

The projection on the sky of the plasma emissivity gives the X--ray
surface brightness: in order to constrain the physical parameters of
the ICM, the observed surface brightness can be either geometrically
deprojected or, more simply, fitted with a model obtained from an
assumed distribution of the gas density.


\subsection{Hydrostatic Models for ICM Distribution}
\label{hydrostatic.sec}
   
From the expression for the sound speed ${c_{\rm s}}^2 = \gamma k
T/\mu m_{\rm p}$, where $\gamma=5/3$ for a monatomic gas, the time
required for a sound wave in the ICM to cross a cluster is given by:
\begin{equation}
\label{1.tempo.suono.eq}
t_{\rm s} \approx 6.6 \times 10^8 \left(\frac{T}{10^8 \mbox{ K}}\right)
^{-\frac{1}{2}} \left(\frac{D}{1 \mbox{ Mpc}}\right) ~~~~~\mbox{ yr}
\end{equation}
where $D$ is the cluster diameter.  Since this time is short compared
to the likely age of a cluster (in first approximation assumed to be
$\sim 10^{10}$ yr), the gas is generally thought to be in hydrostatic
equilibrium in the gravitational potential of the cluster: $ \nabla p
= -\rho \nabla \phi $, where $p = \rho k T/\mu m_{\rm p}$ is the gas
pressure, $\rho $ is the gas density and $\phi$ is the gravitational
potential of the cluster.  Under the assumptions that the ICM is
locally homogeneous and the cluster is spherically symmetric, the
hydrostatic equilibrium equation reduces to:
\begin{equation}
\label{1.idrost.eq}
\frac{1}{\rho } \frac{dp}{dr} = -\frac{d\phi}{dr} = -\frac{G M(r)}{r^2}
\end{equation}
where $r$ is the radius from the cluster center and $M(r)$ is the
total cluster mass within $r$.  If the gas self gravity is ignored,
then the distribution of the ICM is determined by the cluster
potential $\phi(r)$ and the temperature distribution of the gas
$T(r)$, and Eq. \ref{1.idrost.eq} is a linear equation for the
logarithm of the gas density.  Under these assumptions, the
gravitational mass $M_{\rm tot}$ of a galaxy cluster can be written
as:
\begin{equation}
M_{\rm tot}(<r) = - \frac{kT \, r}{G \mu m_{\rm p}}
\left[ \frac{ d \ln \rho}{d \ln r} + \frac{d \ln T}{d \ln r} \right]
\label{mass.eq}
\end{equation}
This expression is commonly used to estimate the gravitational mass of
galaxy clusters and groups from X-ray observations, through the
measurements of the radial profiles of temperature and density
\citep[e.g.,][]{Voigt-Fabian_2006, Gitti_2007c, Gastaldello_2007,
  Ettori_2010}.  However, we note that Eq. \ref{mass.eq} neglects the
contribution of possible additional, non-thermal pressure that, if
present, should be included in the estimate of the total mass. In
particular, recent results from numerical simulations indicate that
the total mass of simulated clusters estimated through the X-ray
approach is lower that the true one due to gas bulk motions
(i.e. deviation from the hydrostatic equilibrium) and the complex
thermal structure of the gas \citep{Rasia_2006, Nagai_2007}.  Possible
observational biases in the derivation of X-ray masses are also
discussed in \citet{Piffaretti-Valdarnini_2008}.


\subsubsection*{The $\beta$--Model}
\label{betamodel.sec}  

\citet{Cavaliere-Fusco_1976} studied the X-ray emission by the hot
plasma in galaxy clusters and developed a hydrostatic model based on
the assumption that the gas and the galaxies are in equilibrium in the
same gravitational potential $\phi$ (see e.g., \citep{Sarazin_1988},
and \citep{Arnaud_2009} for a recent commentary on this model).  By
further assuming that the galaxy distribution is well described by
King's approximation to the isothermal sphere \citep{King_1962}, the
expression for the ICM distribution may be written as
\citep{Cavaliere-Fusco_1976}:

\begin{equation}
\label{1.beta.model.eq}
\rho(r) = \rho_0 \left[1+ \left(\frac{r}{r_{\rm core}}\right)^2\right]^
{-\frac{3}{2} \beta}
\end{equation}
~\\
and the surface brightness profile observed at a projected radius $b$,
$I(b)$, is in the form \citep{Cavaliere-Fusco_1976}:

\begin{equation}
\label{1.beta.bril.eq}
I(b) =  I_0 \left[1+ \left(\frac{b}{r_{\rm core}}\right)^2\right]^
{\frac{1}{2} -3\beta}
\end{equation}
~\\
The parameter $\beta$ is defined as
\begin{equation}
\label{1.beta.eq}
\beta = \frac{{\sigma_{\rm r}}^2}{k T/\mu m_{\rm p}}
\end{equation}
~\\
where $\sigma_{\rm r}$ is the line--of--sight velocity dispersion, and
represents the ratio of specific kinetic energies of galaxies and gas.

This self--consistent isothermal model, called the
\textit{``$\beta$--model''}, is widely used in the X--ray astronomy to
parametrize the gas density profile in clusters of galaxies by fitting
their surface brightness profile.  One of the advantages of using a
$\beta$-model to parameterize the observed X-ray surface brightness is
that the total mass profiles can be recovered analytically and
expressed by a simple formula:
\begin{equation}
M_{tot}(<r) = \frac{k \, r^2}{G \mu m_p} \left[
\frac{ 3 \beta r T}{r^2 + r^2_{\rm c}} - \frac{d T}{d r} \right]
\label{massbeta.eq}
\end{equation}

Eq. \ref{1.beta.model.eq} states that the gas density rises towards
the cluster center.  Since the bremsstrahlung and line emission depend
on the square of the gas density (Eq. \ref{1.em.x.eq}), in the central
regions of clusters the loss of energy by X--ray emission represents
an important process for the thermal particles in the ICM.  In
particular, if the gas density reaches high enough values, large
amounts of gas cool and flow into the centers of clusters, forming the
so--called {\it cooling flows}.  In cooling flow clusters, the single
$\beta$-model is found to be a poor description of the entire surface
brightness profile: a fit to the outer regions shows a strong excess
in the center as compared to the model (see \S \ref{cf.sec}).
Conversely, the centrally peaked emission is a strong indication of a
cooling flow in relaxed cluster.


\subsection{Cool Cores and Cooling Flow Problem}
\label{cf.sec}

The X--rays emitted from clusters of galaxies represent a loss of
energy of the ICM.  The resultant cooling time is calculated as the
time taken for the gas to radiate its enthalpy per unit volume $H_{\rm
  v}$ using the instantaneous cooling rate at any temperature:
\begin{equation}
  t_{\rm cool} \approx \frac{H_{\rm v}}{n_{\rm e} n_{\rm H}
    \Lambda(T)} = \frac{\gamma}{\gamma -1} \frac{kT}{\mu X n_{\rm
      e} \Lambda(T)} \label{tcool.eq}
\end{equation}
where: $\gamma=5/3$ is the adiabatic index; $\mu \approx 0.6$ is the
molecular weight; ${\rm X} \approx 0.71$ is the hydrogen mass
fraction; and $\Lambda(T)$ is the cooling function.  In the central
region, the cooling rate of the ICM is sufficiently high that the
particles lose their energy via radiation, as inferred from X--ray
images of the cores of many clusters which show strongly peaked
surface brightness distributions.  The density of the gas then rises
to maintain the pressure required to support the weight of the
overlying gas in the rest of the cluster, causing a slow subsonic
inflow of material towards the cluster center.  This qualitative
picture describes the physics of the process known as a
\textit{cooling flow} (see \citep{Fabian_1994} for a review of the
standard model, and \citep{Ettori-Brighenti_2008} for a quantitative
description of the evolution of cooling flows).  The steady cooling
flow is confined within the region in which $t_{\rm cool}$ is less
than the time for which the system has been relaxed (thus allowing the
establishment of a cooling flow). This \textit{cooling region} is
delimited by the so--called \textit{cooling radius} $r_{\rm cool}$,
which is usually defined as the radius at which $t_{\rm cool}$ is
equal to the look-back time to $z=1$, i.e. $\sim 7.7 \times 10^9$ yr
in the concordance cosmology.  The fraction of clusters with a central
surface excess with respect to a $\beta$-model, the so-called
\textit{cool cores}, is large.  Cool core clusters are about 90\% of
X-ray-selected clusters with total mass $M_{\rm tot}$ \ltsim $10^{14}
M_{\odot}$, and about 50\% of X-ray-selected clusters with total mass
$M_{\rm tot}$ \gtsim $10^{14} M_{\odot}$ \citep[][]{Chen_2007}.  Cool
cores are also characterized by strong enhancements in the central
abundance \citep[e.g.,][]{DeGrandi-Molendi_2001,
  DeGrandi-Molendi_2009} and declining temperature profiles toward the
central region \citep[e.g.,][]{Allen_2001, Vikhlinin_2005}.

In the standard model, the ``magnitude'' of a cooling flow is measured
from the amount of matter which crosses $r_{\rm cool}$ and flows
towards the center, that is $\dot{M}$, the \textit{mass inflow rate}.
The mass inflow rate, due to cooling, can be estimated from the X--ray
images by using the luminosity $L_{\rm cool}$ associated with the
cooling region and assuming that it is all due to the radiation of the
total thermal energy of the gas plus the $p dV$ work done on the gas
as it enters $r_{\rm c}$: $ L_{\rm cool} = dE/dt $, where $dE =
dE_{\rm th} + p dV = (\gamma/\gamma -1) p \, dV $, and $p dV = (\rho
\, kT \, dV) / (\mu m_{\rm p}) = (dM \, kT)/(\mu m_{\rm p})$, with
$\gamma = 5/3$.  By substituting one obtains the expression for
$L_{\rm cool}$:

\begin{equation}
\label{cf.L.cool.eq}
L_{\rm cool} = \frac{5}{2} \frac{\dot{M}}{\mu m_{\rm p}}k T
\end{equation}
~\\
where $T$ is the temperature of the gas at $r_{\rm cool}$.  $L_{\rm
  cool}$ ranges from $\sim 10^{42}$ to $> 10^{44}$ erg s$^-1$ and
generally represents $\sim 10$\% of the total cluster luminosity
\citep{Fabian_1994}.  Value of $\dot{M} \sim 100 M_{\odot}$ yr$^{-1}$
are fairly typical for cluster cooling flows.

However, the current generation of X-ray satellites, \textit{Chandra}
and \textit{XMM-Newton}, has radically changed our understanding of
cooling flow systems.  Albeit confirming the existence of short
cooling times, high densities and low temperatures in the cluster
cores, the arrival of high-resolution X-ray spectral data has shown
the absence or weakness of the soft X-ray line Fe XVII, indicating
that the amount of gas cooling radiatively below about one third of
its original temperature is ten times less than expected
\citep[e.g.,][]{David_2001, Peterson_2001, Johnstone_2002}.  The lack
of evidence for central gas cooling to very low temperatures at the
rates predicted in the hot atmospheres of galaxy clusters and groups
represents an open question which is often referred to as the so
called {\it 'cooling flow problem'} (see \citep{Peterson-Fabian_2006,
  McNamara-Nulsen_2007, Bohringer-Werner_2010} for reviews).

Historically, two main approaches were adopted to solve this problem.
As the gas radiates but does not appear to cool, either the normal
signatures of radiative cooling below $1-2$ keV are somehow
suppressed, or there must be an energy-injection mechanism into the
ICM which compensates cooling. Different possibilities considered in
the former hypothesis include absorption \citep{Peterson_2001,
  Fabian_2001}, inhomogeneous metallicity \citep{Fabian_2001,
  Morris-Fabian_2003}, and the emerging of the missing X--ray
luminosity in other bands, like ultraviolet, optical and infrared due
to mixing with cooler gas/dust \citep{Fabian_2001, Fabian_2002b,
  Mathews-Brighenti_2003b}.  Proposed heating mechanisms in the
context of the latter approach include e.g., processes associated with
relativistic AGN outflows \citep{Rosner-Tucker_1989,
  Tabor-Binney_1993, Churazov_2001, Bruggen-Kaiser_2001,
  Kaiser-Binney_2003, Ruszkowski-Begelman_2002,
  Brighenti-Mathews_2003}, electron thermal conduction from the outer
regions of clusters \citep{Tucker-Rosner_1983, Voigt_2002,
  Fabian_2002c, Zakamska-Narayan_2003}, continuous subcluster merging
\citep{Markevitch_2001}, contribution of the gravitational potential
of the cluster core \citep{Fabian_2003a}, feedback from intra-cluster
supernovae \citep{Domainko_2004}.  Among all these, feedback by the
central AGN appears to be the most promising solution.


\section{X-ray cavities and shocks: AGN feedback in action}
\label{feedback.sec}  

It was already known in the early 90s that central dominant (cD)
galaxies of cool core clusters have a high incidence of radio
activity, showing the presence of central FR-I radiogalaxies in 70\%
of the cases \citep{Burns_1990, Best_2007}.  Their behaviour differs
from that of quasar: in many low-accretion-rate AGNs almost all the
released energy is channelled into jets because the density of the gas
surrounding the black hole is not high enough for an efficient
radiation \citep[e.g.,][]{Churazov_2005}. In fact, the importance of
these objects has been underestimated for a long time due to their
poor optical luminosity.  The importance of radio galaxies in cool
cores began to emerge after the discovery, with the X-ray satellite
{\it ROSAT}, of deficits in the X-ray emission of the Perseus and
Cygnus A clusters which are spatially coincident with regions of
enhanced synchrotron emission \citep{Bohringer_1993, Carilli_1994}.
With the advent of the new high-resolution X-ray observations
performed with the current generations of X-ray telescopes, {\it
  Chandra} and {\it XMM-Newton}, it became clear that the central
radio sources have a profound, persistent effect on the ICM.  In
particular, \textit{Chandra} images, which are obtained at the superb
angular resolution of $\sim$1$''$, showed that the Perseus and Cygnus
A clusters are not isolated cases -- indeed the central hot gas in
many cool core systems is not smoothly distributed, but is instead
cavitated on scales often approximately coincident with lobes of
extended radio emission.  These observations also reveal highly
disturbed structures in the cores of many clusters, including shocks,
ripples and sharp density discontinuities. The comparison with radio
images obtained at similar angular resolution has revealed that AGN
jets are the cause of many of these disturbances.  The most typical
configuration is for jets from the central dominant elliptical of a
cluster to extend outwards in a bipolar flow, inflating lobes of
radio-emitting plasma. These lobes push aside the X-ray emitting gas
of the cluster atmosphere, thus excavating depressions in the ICM
which are detectable as apparent {\it 'cavities'} in the X-ray images.
The brightness depressions observed in X-rays, which are mostly found
in pairs, are $\sim 20-40$\% below the level of the surrounding gas,
consistently with the expected decrement along an empty bubble
immersed in a $\beta$-model atmosphere \citep{Fabian_2000,
  McNamara_2000, Blanton_2001, Blanton_2003, Nulsen_2002}.  The
cavities are often surrounded by bright shells, or rims, which are
typically found to be cooler than the ambient medium
\citep{Fabian_2000, McNamara_2000, Blanton_2001, Blanton_2003,
  Doria_2011}. This is likely due to the compression of the central,
low-entropy gas into the bright shell during the cavity rising and
expansion into the hot atmosphere.

X-ray cavities are present in \gtsim 70\% of cool core clusters
\citep{Dunn-Fabian_2006}, but this fraction could be underestimated
due to the limitation of cavity detectability \citep{Birzan_2009}.
Identifying radio galaxies as a primary source of feedback in the hot
atmospheres of galaxy clusters and groups has been one of the major
achievements of the current generation of X-ray observatories (for a
comprehensive review see \citep{McNamara-Nulsen_2007} and references
therein).  Well-studied examples of cavity clusters are:
Perseus \citep[e.g.,][]{Bohringer_1993, Churazov_2000, Fabian_2000,
  Fabian_2006},
%
Hydra A \citep[e.g.,][]{McNamara_2000, David_2001, Nulsen_2005b,
  Wise_2007, Simionescu_2009b, Kirkpatrick_2009b, Gitti_2011a},
%
M 87 \citep[e.g.,][]{Bohringer_1995, Churazov_2001, Molendi_2002,
  Forman_2007, Simionescu_2008, Million_2010},
%
A 2052 \citep[e.g.,][]{Blanton_2001, Blanton_2004, Blanton_2009,
  Blanton_2011},
%
%
RBS 797 \citep[e.g.,][]{Schindler_2001, Gitti_2006, Cavagnolo_2011,
  Doria_2011},
%
%
%
%
%
A 133 \citep[e.g.,][]{Fujita_2002, Randall_2010},
%
%
%
A 262  \citep[e.g.,][]{Blanton_2004, Clarke_2009},
%
MS 0735+7421  \citep[e.g.,][]{McNamara_2005, Gitti_2007a}.
%
In-depth analyses of individual objects, which are now numerous in the
literature, combined with studies of cavity samples
\citep{Birzan_2004, Birzan_2008, Dunn_2005, Dunn-Fabian_2006,
  Rafferty_2006, Diehl_2008} have enabled us to identify the global
properties which are common among the cavities, thus shedding light of
the feedback mechanism.  The emerging picture is that bipolar outflows
emanating from the BCG core inflate large bubbles while driving weak
shocks, heat the ICM and induce a circulation of gas and metals on
scales of several 100s kpc. Weak shocks have been observed as
ripple-like features in the ICM in the deepest X-ray images of Perseus
and A 2052 \citep{Fabian_2006, Blanton_2011}.

However, the differences between groups and clusters imply that the
existing studies on cavities in clusters tell us little about how
feedback operates in groups.  With respect to rich galaxy clusters,
the observation of cavities in galaxy groups and ellipticals is
complicated by the lower X-ray surface brightness, which limit their
detection in shallow X-ray images.  On the other hand, there are
several examples of AGN-ICM interaction just a few tens Mpc away from
us which allow us to probe regions closer to the central black hole.
In particular, low mass systems with cavities which now have deep {\it
  Chandra} images are:
M 84 \citep{Harris_2002, Finoguenov_2008},
NGC 4636 \citep{Jones_2002, O'Sullivan_2005, Baldi_2009},
NGC 5044 \citep{Buote_2003a, Gastaldello_2009, David_2009,
  David_2011},
HCG 62 \citep{Vrtilek_2002, Morita_2006, Gu_2007, Gitti_2010},
NGC 5846 \citep{Machacek_2011}.
Performing in-depth individual studies and sample studies of the
lower-energy outbursts in these smaller systems is of major interest
because the relationship between AGNs and hot gas can significantly
influence galaxy evolution in the group environment, which is the
locus of the majority of galaxies in the Universe \citep{Eke_2004}.
Due to the shallower group potential, the AGN outburst can have a
large impact on the intra-group medium in terms of altering the
thermal history and spatial distribution of the intra-group medium, as
the mechanical output by radio AGN is of the same order of magnitude
as the binding energy of groups \citep{Giodini_2010}.  Such
investigations have been undertaken only recently for individual
objects (e.g., NGC 5813: \citep{Randall_2011}, AWM 4:
\citep{O'Sullivan_2010, O'Sullivan_2011a}) and for group samples
\citep{Johnson_2009, Sun_2009b, Dong_2010, Giacintucci_2011b}, and are
rapidly improving our understanding of these systems. However, this
observational effort is still awaiting detailed theoretical work in
order to corroborate the observational findings. Recent detailed
simulations indicate that groups are not simply a scaled-down version
of clusters, as there may be remarkable differences between how AGN
feedback operates in galaxy group and in galaxy cluster environments
\citep{Gaspari_2011b}.  In particular, AGN heating in groups seems to
act through persistent, gentle injection of mechanical energy.  On the
other hand, in clusters there must be also the action of rare,
powerful outbursts \citep{Gaspari_2011b}, although more extensive
theoretical work is required to reach firm conclusions.

\begin{figure*}[ht]
\centerline{
\includegraphics[width=8.1cm]{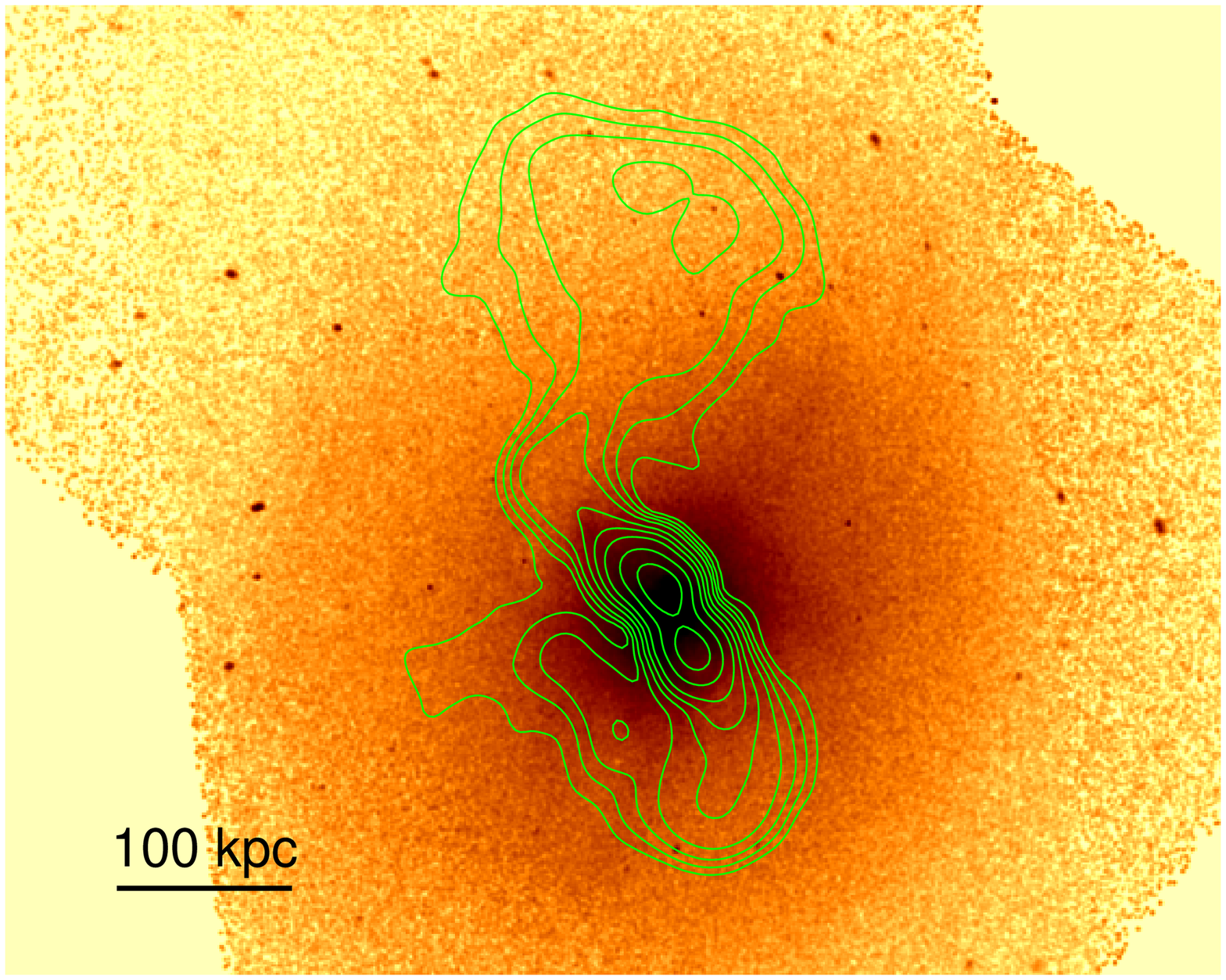}
\includegraphics[width=8.1cm]{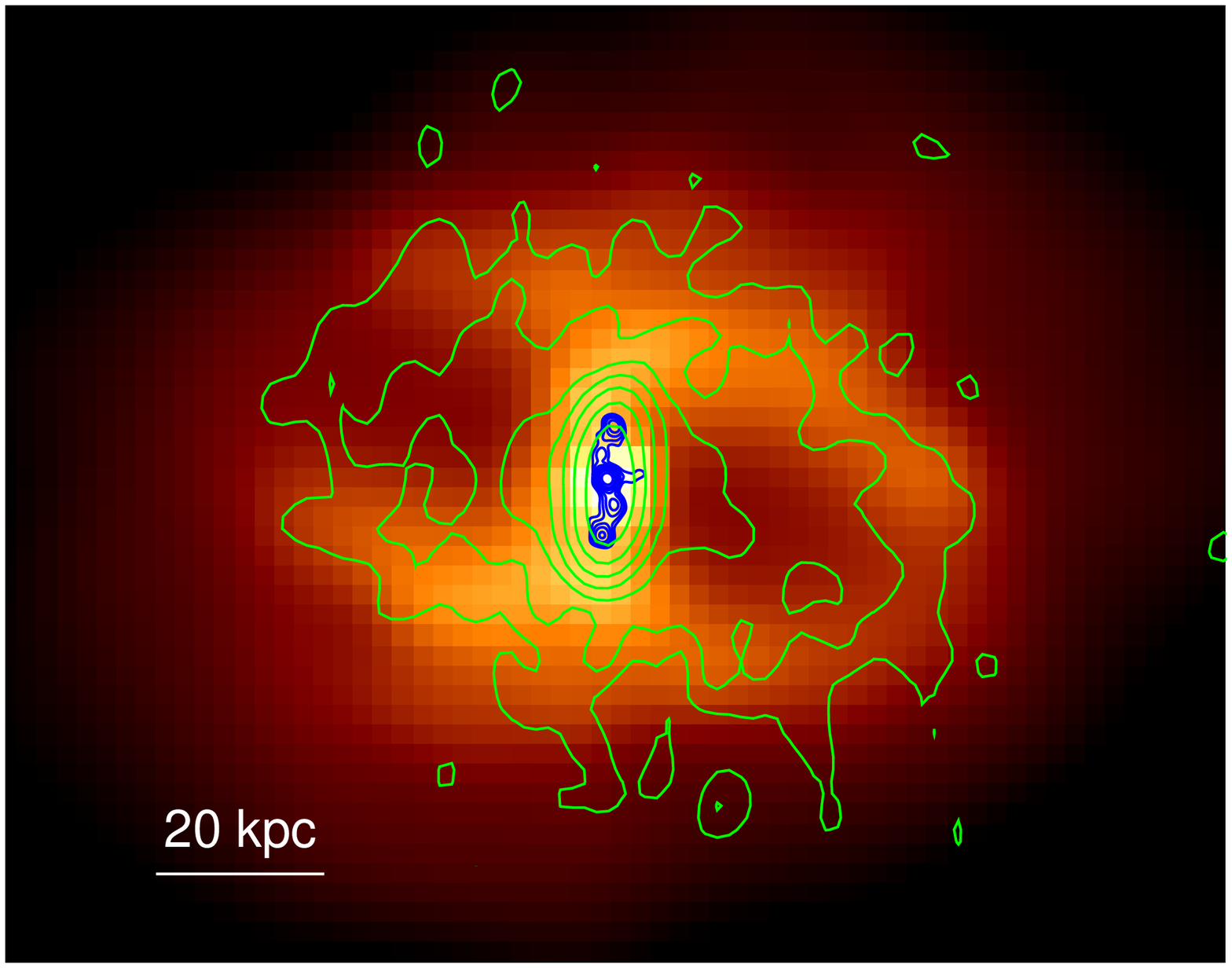}
}
\centerline{
\includegraphics[width=8.1cm]{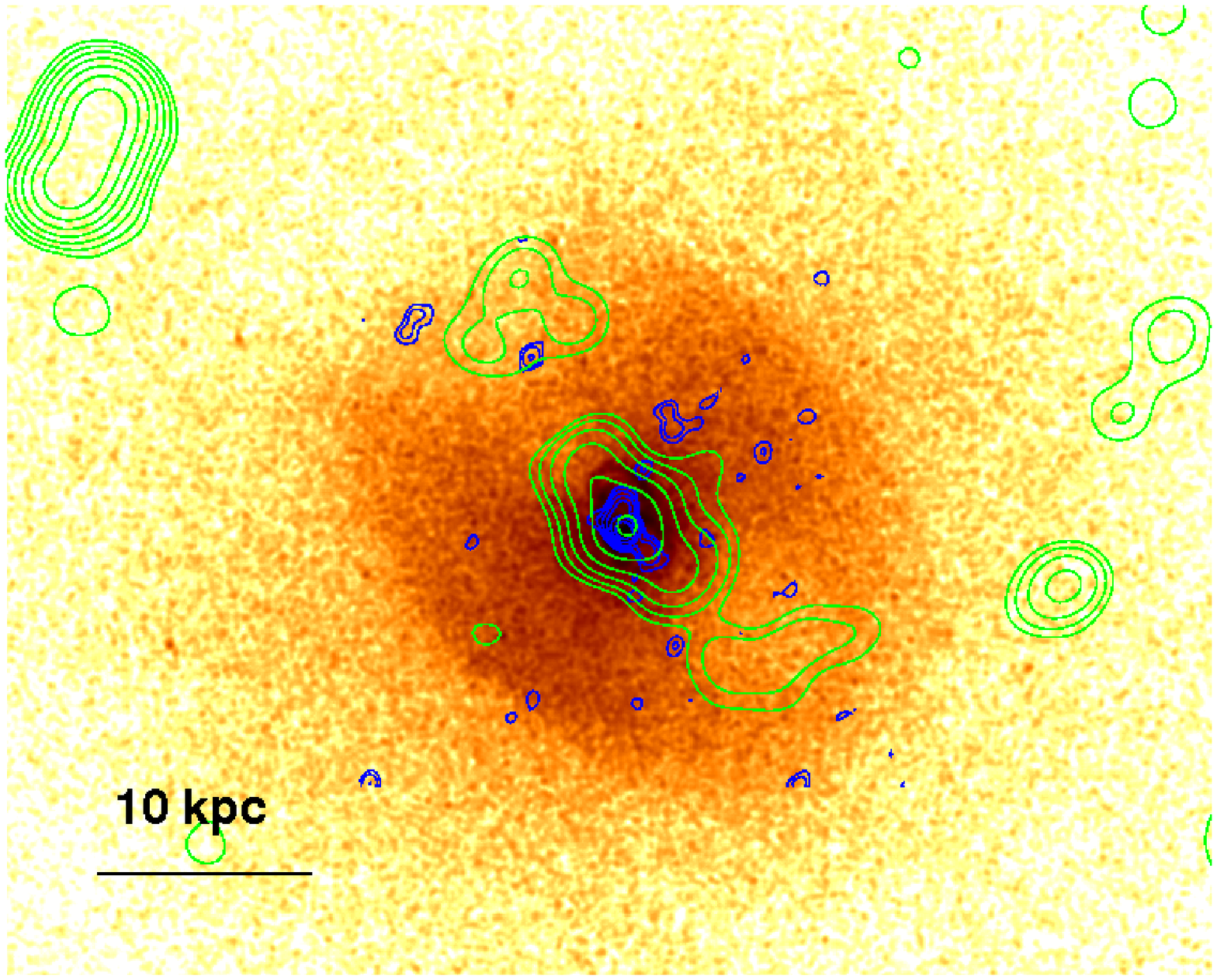}
\includegraphics[width=8.1cm]{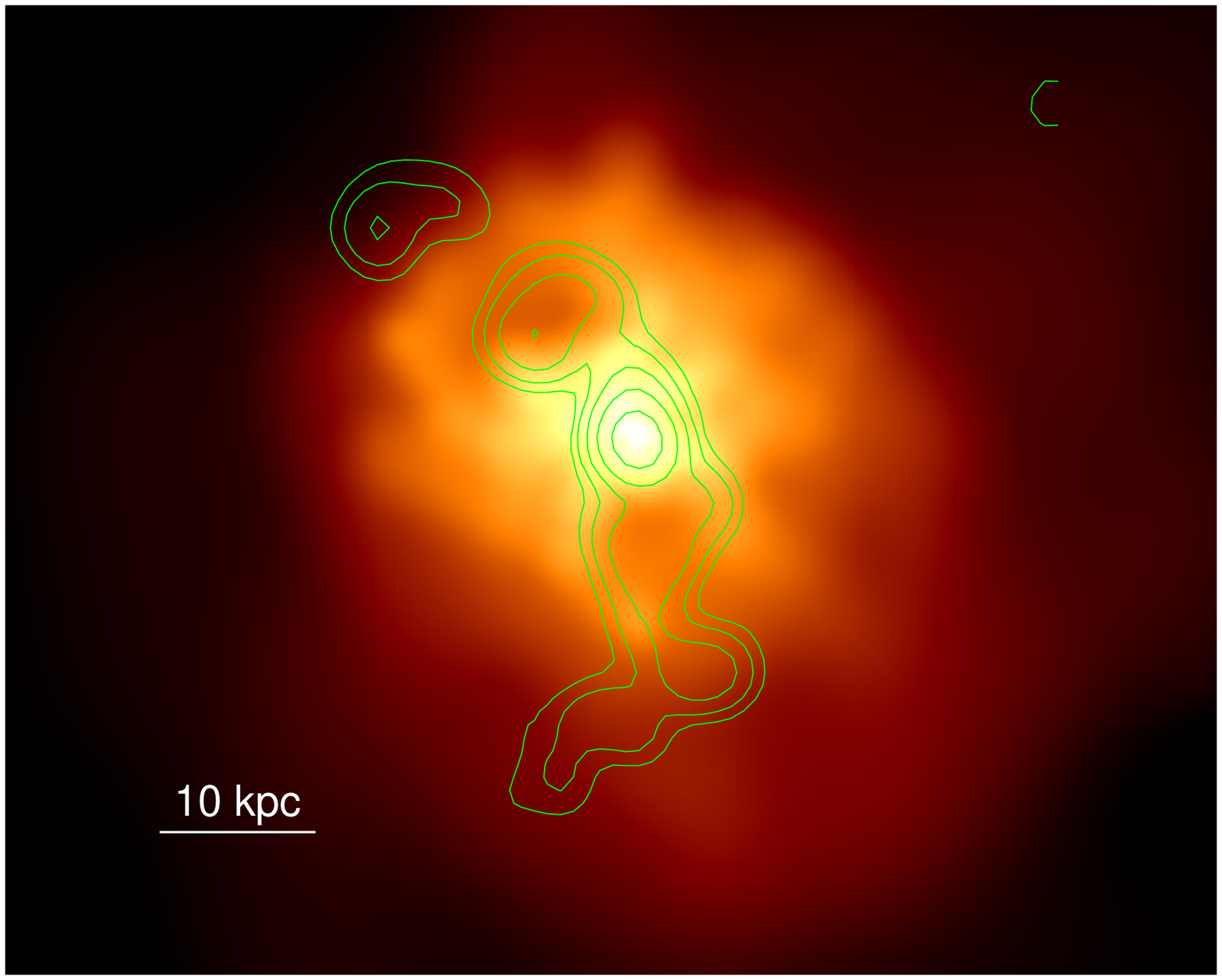}
}
\caption{\small \label{examples.fig} \textit{Top Left:} The green
  contours outlining the 330 MHz radio emission from \citet{Lane_2004}
  are overlaid onto the 0.5--7.5 keV {\it Chandra} image of the galaxy
  cluster {\bf Hydra A} (z=0.0538).  The extended radio lobes fill a
  large-scale system of X-ray cavities and are surrounded by a
  ``cocoon'' shock. See also \S \ref{giant.sec} and \S
  \ref{mechanical.sec}.  Adapted from \citet{Nulsen_2005b}.
  \textit{Top Right:} Very Large Array (VLA) radio contours overlaid
  onto the 0.5--7.0 \textit{Chandra} image of the galaxy cluster {\bf
    RBS 797} (z=0.35).  The subarsec resolution radio image shows the
  details of the innermost 4.8 GHz radio jets (blue contours), which
  clearly point in a north-south direction.  Remarkably, these inner
  jets are almost perpendicular to the axis of the 1.4 GHz emission
  observed at $1''$ resolution (green contours), which is elongated in
  the northeast-southwest direction filling the X-ray cavities.
  Adapted from \citet{Gitti_2006}.  \textit{Bottom Left:} 0.3--2 keV
  {\it Chandra} image of the galaxy group {\bf NGC 5813} (z=0.0066)
  with 1.4~GHz VLA (blue) and 235~MHz Giant Metrewave Radio Telescope
  (green) radio contours overlaid. The image shows two pairs of
  cavities, plus an outer cavity to the northeast, two sharp edges to
  the northwest and southeast, and bright rims around the pair of
  inner cavities. Adapted from \citet{Randall_2011}.  \textit{Bottom
    Right:} 235 MHz GMRT contours overlaid on the smoothed 0.5--2.0
  keV {\it Chandra} image of the compact group {\bf HCG 62}
  (z=0.0137).  The radio source shows a disturbed morphology with
  inner lobes clearly filling the well defined X-ray cavities, but
  with outer lobes having no associated X-ray cavities (see also \S
  \ref{lobes.sec}).  Adapted from \citet{Gitti_2010}.  }
\end{figure*}

Examples of well-studied cavity systems in clusters and groups are
shown in Figure \ref{examples.fig}.


\subsection{Cavity Heating}
\label{heating.sec}

The heating is thought to occur through the dissipation of the cavity
enthalpy and through shocks driven by the AGN outburst.  The energy
required to create a cavity with pressure $p$ and volume $V$ is the
sum of the $pV$ work done by the jet to displace the X-ray emitting
gas while it inflates the radio bubble, and the internal energy of the
lobes, i.e. the enthalpy given by:

\begin{equation}
\label{ecav.eq}
E_{\rm cav} \equiv H = E_{\rm int} + pV = \frac{\gamma}{\gamma - 1} pV
\end{equation}

where $\gamma$ is the ratio of the specific heats of the cavity
content.  If the lobes are dominated by the magnetic field, by
non-relativistic gas, or by relativistic plasma, $H$ can vary from
$2pV$ and $4pV$.  In particular, typically it is assumed that the
internal composition of the cavity is dominated by relativistic
plasma, therefore $\gamma = 4/3$ and $H = 4 pV$.  The product of
pressure and volume can be estimated directly by X-ray observations
through measurements of the cavity size and of the temperature and
density of the surrounding ICM.
A potential issue is represented by the uncertainties in the
measurement of the cavity volume. The cavity size is usually estimated
through a visual inspection of the X-ray images. This method is
therefore dependent on the quality of the X-ray data, and also subject
to the arbitrariness of the observer.  The cavity size and geometry
measured by different observers may vary significantly depending on
the approach adopted, leading to differences between estimates of up
to a factor of two in $pV$ \citep[e.g.,][]{Gitti_2010,
  O'Sullivan_2011b, Cavagnolo_2011}.

Systematic observational studies of samples of X-ray cavities show
that their enthalpies measured from Eq. \ref{ecav.eq} lie between
$\sim 10^{55}$ erg (in ellipticals, groups and poor clusters) and
\gtsim $10^{61}$ erg (in rich clusters).  On the other hand,
simulations indicate that $pV$ varies with time during the cavity
evolution and may be an inaccurate measure of the total energy
released \citep{Mathews-Brighenti_2008a,
  Mathews-Brighenti_2008b}. Cavity power estimates within a factor of
two of the simulated values seem possible provided the inclination
angle of the jets is known accurately \citep{Mendygral_2011}.  Bearing
this caveat in mind, when divided by the cavity age, $t_{\rm cav}$,
the observational measurements give an estimate of the so called
``cavity power'', $P_{\rm cav}$.  Since shocks are very difficult to
detect and are currently known only in a few systems (e.g., Hydra A
\citep{Nulsen_2005b}, MS 0735+7421 \citep{McNamara_2005}, HCG 62
\citep{Gitti_2010}, NGC 5813 \citep{Randall_2011}), for consistency
the usual approach in sample studies is that of considering only the
cavity power. $P_{\rm cav}$ thus provides a lower limit (and
best-available gauge) to the true total mechanical power of the AGN,
i.e. the jet power: $P_{\rm jet}$ \gtsim $P_{\rm cav} = E_{\rm cav} /
t_{\rm cav}$.

As proposed by \citet{Birzan_2004}, the cavity age can be estimated in
three ways: (1) by assuming that the cavity rises the hot gas
atmosphere at the sound speed ${c_{\rm s}} = \sqrt{\gamma k T/\mu
  m_{\rm p}}$: in this case the cavity reaches the projected distance
$R$ from the cluster center in the sound crossing time $t_s = R/c_s =
R / \sqrt{\gamma kT / \mu m_{\rm p}}$; (2) by assuming that the cavity
is buoyant and move outwards at the terminal velocity $v_{\rm t} =
\sqrt{2 g V / S C}$, where $g = G M_{< R} / R^2$ is the gravitational
acceleration at the cavity position $R$, $V$ is the volume of the
cavity, $S$ is the cross-section of the cavity, and $C=0.75$ is the
drag coefficient \citep{Churazov_2001}: in this case the cavity age is
the buoyancy-time $t_{\rm buoy} \sim R / \sqrt{2 g V / S C}$ ; (3) by
considering the time required for gas to refill the displaced volume
of the cavity as it rises: in this case the cavity age is estimated as
$t_{\rm ref} \sim 2 \sqrt{r / g}$, where $r$ is the radius of the
cavity.  Typically the age estimates agree to within a factor of 2,
with the buoyancy times lying in between the sound crossing time and
the refill times. Most sample studies adopt the buoyancy time, which
for typical values gives cavity ages of the order of few $10^7$ yr
\citep[e.g.,][]{Rafferty_2006}.


\subsection{The Relationship between Jet Power and $L_{cool}$}
\label{pcav.sec}

\begin{figure*}
\vspace{0cm}
\centerline{
\includegraphics[width=8.3cm]{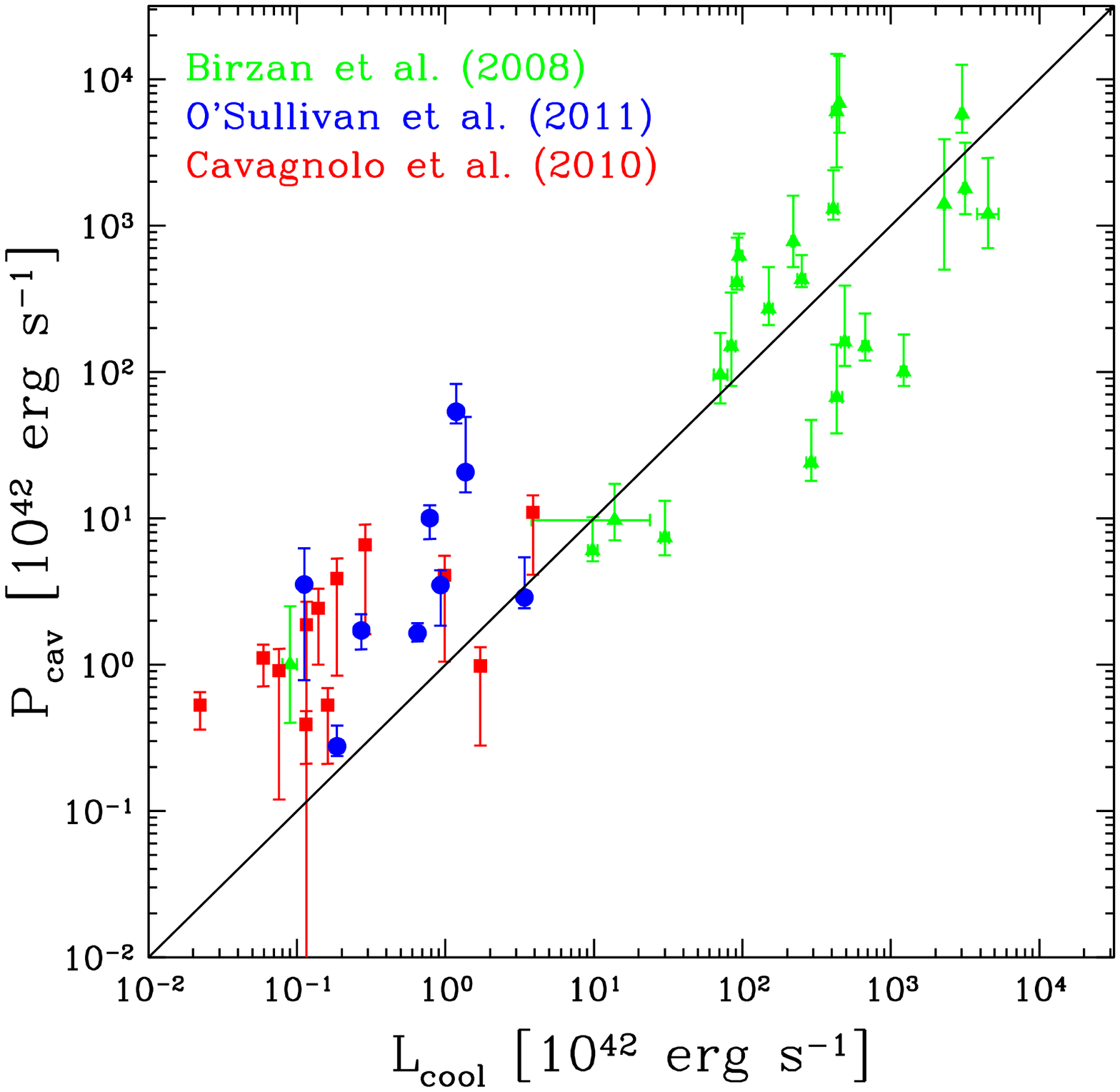}
\includegraphics[width=8.3cm]{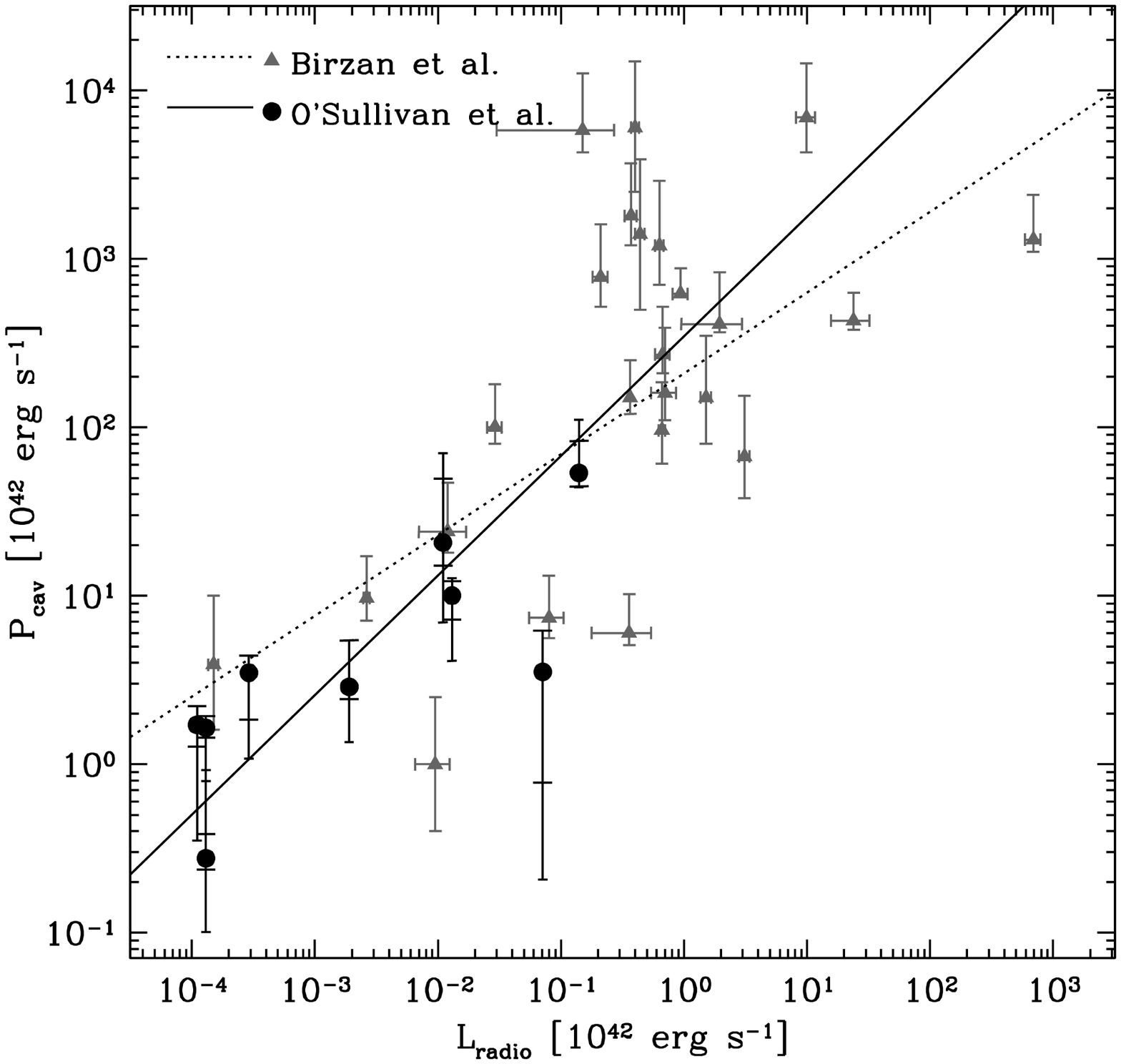}
}
\vspace{-2.5cm}
\caption{\small \label{Pcav.fig} {\it Left:} Cavity power of the
  central AGN, $P_{\rm cav}$, versus the X-ray luminosity of the ICM
  inside the cooling region, $L_{\rm cool}$. The cavity power is
  calculated assuming $4 pV$ of energy per cavity and the buoyancy
  timescale.  Different symbols denotes systems in different samples
  presented in the literature: {\it green triangles} --
  \citet{Birzan_2008}, {\it red squares} -- \citet{Cavagnolo_2010},
  {\it blue circles} -- \citet{O'Sullivan_2011b}.  1$\sigma$
  uncertainties on cavity power are indicated by error bars (see
  \citep{O'Sullivan_2011b} for details).  The diagonal line denotes
  $P_{\rm cav} = L_{\rm cool}$. Credit O'Sullivan (private
  communication).  {\it Right:} Cavity power of the central AGN,
  $P_{\rm cav}$, versus integrated 10~MHz-10~GHz radio power, $L_{\rm
    radio}$, for the systems in the sample of B\^{i}rzan et al. (grey
  triangles) and the groups in the sample of O'Sullivan et al. (black
  circles).  The solid fit line indicates the regression fit to the
  data points calculated by \citet{O'Sullivan_2011b}. The dotted line
  indicates the relation found by \citet{Birzan_2008}.  Adapted from
  \citet{O'Sullivan_2011b}.}
\end{figure*}

Once a cavity is detected, it is relatively simple to estimate its
power from the measurements of $E_{\rm cav}$ and $t_{\rm cav}$ by
applying Eq. \ref{ecav.eq}.  The cavity power, $P_{\rm cav}$, which is
a measure of the energy injected into the hot gas by the AGN outburst,
can then be compared directly with the gas luminosity inside the
cooling radius, $L_{\rm cool}$, which represents the luminosity that
must be compensated for by heating to prevent cooling.  The gas
luminosity inside the cooling radius is estimated as the bolometric
X-ray luminosity derived from a deprojection spectral analysis.  In
Figure \ref{Pcav.fig} (left panel) is shown a quantitative comparison
between $P_{\rm cav} = 4 pV / t_{\rm buoy}$ and $L_{\rm cool}$
calculated for the extended sample discussed in
\citet{O'Sullivan_2011b}, who combined new data of 9 galaxy groups
with the cluster sample of \citet{Birzan_2008} and with the elliptical
sample of \citet{Cavagnolo_2010}.

This plot follows those presented in Figure 2 of \citet{Birzan_2004}
and in Figure 6 of \citet{Rafferty_2006}. As already noted by these
authors, it is evident that the cavity power scales in proportion to
the cooling X-ray luminosity, although with a big scatter.  In
general, it appears that the high mass (corresponding to high
$L_{cool}$) systems need an average of $4 pV$ per cavity to counter
cooling. On the other hand, if we recalculate $P_{\rm cav}$ as $1 pV /
t_{\rm buoy}$ all the points in the plot will shift down by a factor
4, and only the lower mass systems will still lie around the line
$P_{\rm cav} = L_{\rm cool}$. These systems require $1 pV$ per cavity
to offset cooling at the present time.
A few low mass systems will even still be above the equality line,
thus indicating that the total mechanical power of the AGN far exceeds
the radiative losses and their atmospheres are being heated.
Although this extended sample is not a complete sample and therefore
is not representative of the whole population of clusters, groups and
ellpticals, it is interesting to consider the mean values of heating
and cooling to see how they compare.  We estimated a mean cooling
power of $4.09 \times 10^{44}$ erg s$^{-1}$, and a mean cavity power
of $6.18 \times 10^{44}$ erg s$^{-1}$.
In order to quantify properly the contribution of AGN feedback, over
the system lifetime, in the energetics of cooling flow, it is
important to determine the ``duty-cycle'' of AGN.  Many studies have
attempted such calculation by adopting different approaches, e.g., by
considering the luminosity function of radio galaxies
\citep{Nipoti-Binney_2005}, the fraction of clusters that contain
bubbles and cavities \citep{Dunn_2005, Dunn-Fabian_2006,
  Dunn-Fabian_2008}, the frequency of bubble required to produce
sufficient heating \citep{Pope_2006}, the prevalence of radio-loud AGN
\citep{Best_2007, Dunn_2010}.  In particular, by considering the
cavities as tracers of the feedback mechanism, i.e. by assuming that
the feedback is active and efficient as long as the cavities are
visible, we can correct the mean cavity power by the fraction of cool
core clusters with cavities, estimated by \citet{Dunn-Fabian_2006} of
the order of at least 70\%. The ratio of mean cavity power to cooling
flow power is thus very close to unity.  The mean values for the whole
sample are only indicative and do not reflect the different behaviour
of groups and ellipticals with respect to clusters.
In particular, such mean powers miss the point that in order to
counter cooling the low mass systems require outbursts with relatively
less total energy, lower powers and repeating more rapidly than high
mass systems.  This is supported by recent numerical simulations of
galaxy groups which show that, in contrast to galaxy clusters, the AGN
self-regulated feedback must act through a quasi-continuous gentle
injection with subrelativistic outflows, rather than through rare and
powerful episodes \citep{Gaspari_2011b}.
An attempt to produce more meaningful averages could be that of
dividing the sample plotted in Figure \ref{Pcav.fig} (left panel) in
two subsets. In fact, by doing this we find that the ratio of mean
cavity power to cooling power for the groups and ellipticals is 7.94
(samples of \citet{Cavagnolo_2010} and \citet{O'Sullivan_2011b}),
compared to a ratio of 1.51 calculated for clusters only (sample of
\citet{Birzan_2008}). If the duty-cycle of low mass systems is the
same as (or not lower than) high mass systems \citep{Dunn-Fabian_2006,
  Dunn_2010}, the relative ratio of heating to cooling appears to be a
factor \gtsim 5 higher in low mass systems.
In other words, groups and ellipticals seem to have five times as much
power available to counter cooling than rich clusters.

A study of a complete, unbiased sample including both cool core and
non-cool core systems is necessary to derive definite constraints on
the balance between heating and cooling. However, it seems plausible
that the time-averaged AGN feedback balances radiation losses of the
ICM.  Therefore the general picture emerging from the observed trend
between X-ray luminosity and bubble mechanical luminosity, together
with the existence of short central cooling time, is that the AGN is
fueled by a cooling flow that is itself regulated by feedback from the
AGN.  The basic idea of this AGN-cooling flow scenario is that a
self-regulated equilibrium may be achieved, in which the radiative
losses from the thermal ICM are balanced by mechanical heating from
the central AGN over the system lifetime.  Although this scenario is
no longer in doubt, is is still not clear how heating can act
preserving at the same time the observed temperature gradient and the
cool core \citep[e.g.,][]{Brighenti-Mathews_2002}.

\subsection{The Relationship between Jet Power and Radio Power}
\label{pjet.sec}

Studies of cavity samples allow to derive the relationship between the
mechanical power and radio emission of AGN jets and lobes. Such a
relationship is of great interest because it helps to understand the
physics of AGN jets \citep[e.g.,][]{Willott_1999, O'Sullivan_2011b},
and because it provides an estimate of the energy available from AGN
based directly on the radio data \citep[e.g.,][]{Best_2007}, thus
avoiding the problem of cavity detectability in shallow X-ray images.
\citet{Birzan_2008} studied a sample dominated by galaxy clusters and
derived the relation between cavity power and 327 MHz radio power, as
well as between cavity power and the integrated 10 MHz - 10 GHz radio
luminosity, extending to lower frequencies their previous work
\citep{Birzan_2004}.  While many of the observed X-ray cavities are
filled with 1.4 GHz radio-emitting plasma, some are undetected at this
high frequency and have been referred to as "ghost cavities". These
may result from the aging of the relativistic particle population or
may have been inflated by events which produced only particles of low
energy.  Examinations of radio images at multiple (and low)
frequencies is particularly important as the progressive loss of
particle energy causes higher frequency emission to fade fastest and
the spectral index to steepen, so that evidence of a former AGN
activity may be reflected only at low frequencies.  The lack of 1.4
GHz radio emission is observed more frequently in groups than in
clusters, therefore low-frequency radio observations are crucial for
galaxy groups.

\citet{Giacintucci_2011b} selected a compilation of 18 galaxy groups,
based on the presence of signs of interaction between the hot gas and
the central AGN, and observed both by the Giant Metrewave Radio
Telescope (GMRT) at frequencies $\leq$ 610 MHz and by {\it Chandra} and/or
{\it XMM-Newton}. These authors found that nine of these groups have
cavities clearly correlated with radio structures. By adding such
systems to the \citet{Birzan_2008} sample, \citet{O'Sullivan_2011b}
examined the relations between jet mechanical power and radio power in
a combined sample which includes the groups having the most reliable
radio measurements currently available.  In particular, the integrated
10 MHz - 10 GHz radio luminosity estimated from the source spectral
index is considered by these authors as a superior cavity power
indicator compared to estimates at a single frequency, since it
accounts for variations in spectral index between sources.  Figure
\ref{Pcav.fig} (right panel) shows the relationship between cavity
power, P$_{\rm cav}$, and the integrated radio luminosity, L$_{\rm
  radio}$, for the combined sample.  The best fitting power-law
relationship is \citep{O'Sullivan_2011b}:

\begin{equation} 
{\rm log}\; {\rm P}_{\rm cav} = 0.71\; (\pm0.11)\;
{\rm log}\; {\rm L}_{\rm radio} + 2.54\; (\pm0.21)
\end{equation}

where P$_{\rm cav}$ and L$_{\rm radio}$ are in units of 10$^{42}$ erg
s$^{-1}$.  See  \citet{O'Sullivan_2011b} for a detailed discussion 
of this relation.


\subsection{Radio Lobe Composition}
\label{lobes.sec}

When the radio source is filling the cavities, it is possible to
compare the radio pressure of the relativistic plasma internal to the
lobes with the X-ray pressure of the surrounding thermal gas.  The
pressure of the hot gas is measured from the density and temperature
derived from the X-ray data as $p \simeq 2 n_{\rm e} k T $.  The total
pressure in a radio lobe is the sum of the magnetic pressure, $p_{B}$,
and the total particle pressure, $p_{\rm part}$, and can be written as
\begin{equation}
  p_{\rm radio} = p_{B} + p_{\rm part} 
= \frac{B^2}{8 \pi} + \frac{1}{3} \frac{E_{\rm part}}{f \, V} 
= \frac{B^2}{8 \pi} + \frac{1}{3} \frac{(1+k) E_{\rm e}}{f \, V}
\label{pradio.eq}
\end{equation}
where $B$ is the magnetic field, $k$ is the ratio of the energy in
protons to that in electrons ($E_{\rm e}$), $V$ is the volume of the
radio lobe and $f$ is the volume filling factor of the relativistic
plasma.  Using the expression for $E_{\rm e}$ given in
\citet{Pacholczyk_1970}, Eq. \ref{pradio.eq} determines the lobe
pressure in terms of the magnetic field strength and the factor $k/f$,
once the volume $V$ of the radio lobe is known.  This calculation is
usually performed under the widely adopted minimum energy conditions,
in which the relativistic plasma is in equipartition with the magnetic
field ($B_{\rm eq}$).  Further assumptions usually made in literature
are $f=1$ and $k=0$ or $k=1$.  A volume filling factor of 1 indicates
that the lobes are empty of thermal gas, which is a reasonable
hypothesis when they are observed to be spatially coincident with
X-ray cavities.  The assumption $k=1$ implies that half of the energy
in particles is in the form of non-radiating particles, as in an
electron-proton jet, whereas $k=0$ would indicate an electron-positron
jet.  We stress that the uncertainties in the calculation of $B_{\rm
  eq}$ and $p_{\rm radio}$ come from the values of $k$ and $f$, which
are still largely unknown. Conversely, it is possible to constrain the
ratio $k/f$ by assuming pressure balance (see below).

For historical reasons the frequency band adopted to calculate the
standard equipartition field is $\nu_1=10$ MHz - $\nu_2=100$ GHz,
i.e. roughly the frequency range observable with radio telescopes.
From a physical point of view, the adoption of this frequency band in
the calculation of the minimum energy is equivalent to the assumption
that only electrons emitting between 10 MHz - 100 GHz, i.e. with
energy between $\gamma_{\rm min} \propto (\nu_1/B_{\rm eq})^{1/2}$ and
$\gamma_{\rm max} \propto (\nu_2/B_{\rm eq})^{1/2}$ are present in the
radio source.  This approach neglects the contribution of the
electrons emitting below 10 MHz and, as a more serious bias, in radio
sources with different $B_{\rm eq} $ selects different energy bands of
the electron population because the energy of the electrons which emit
synchrotron radiation at a given frequency depends on the magnetic
field intensity \citep{Brunetti_2002}.  A more consistent approach is
to calculate the minimum energy conditions, in which $B_{\rm eq}$ does
not depend on the emitted frequency band but directly on the low
energy cut-off of the electron spectrum (typically assumed to be
$\gamma_{\rm min} = 100$). These so-called {\it ``revised''}
equipartition conditions select also the contribution to the
energetics due to the low-energy electrons \citep{Brunetti_1997}.

It is typically found that in cavity systems the X-ray pressure is
more than one order of magnitude higher than the radio pressure
\citep[e.g.,][]{Blanton_2001, DeYoung_2006, Croston_2008, Gitti_2010}.
It is also found that with revised equipartition the cavities are
closer to pressure balance than they are with standard equipartition
\citep[e.g.,][]{Gitti_2010,O'Sullivan_2011b}.  Vice versa, by assuming
the lobes are in pressure equilibrium with the ambient gas it is
possible to constrain the particle content within the radio lobes
\citep{Dunn-Fabian_2004, DeYoung_2006, Birzan_2008}.  In particular,
one can determine the ratio $k_{\rm bal}/f$ that is required to
achieve pressure balance under revised equipartition conditions.
Several studies of the energetics and particle content of the radio
lobes in cooling cores have found high values of $k_{\rm bal}/f$, up
to several thousands (with standard equipartition) for active bubbles
\citep[e.g.,][]{Dunn_2005, Birzan_2008}, suggesting that a large
fraction of energy must be in non-radiating particles if $f$ is close
to unity.  On the other hand, the pressure imbalance found in the
lobes of FR-I radio galaxies in a sample of galaxy groups appears to
be linked to the radio-source morphology, i.e. 'plumed' sources
typically have larger pressure deficits than 'bridged' sources where
the jets are embedded in the lobes \citep{Croston_2008}.  The authors
interpret this result as evidence that plumed sources have a higher
entrainment rate due to the larger fraction of the jet surface which
is in direct contact with the external medium, leading to an increase
in $k/f$.  Although the classification into bridged and plumed
morphologies may not directly apply to radio sources at the center of
cool core systems, typically having amorphous structures, this picture
is consistent with the results of \citet{Dunn_2006} who argue that the
large pressure imbalance observed in radio bubbles as those of the
Perseus cluster is more likely to be due to entrainment rather than a
relativistic proton population. Recent studies show lobes having no
associated X-ray cavities
\citep[e.g.,][]{Gitti_2010,O'Sullivan_2011b}.  Assuming their
detection is not limited by the sensitivity of the current {\it
  Chandra} images, this suggests the possibility of mixing between
ambient gas and radio plasma in the lobes. Therefore the $k_{\rm
  bal}/f > 0$ values measured in such lobes is likely the results of
entrainment of thermal gas through the hot gas atmosphere rather than
an evidence of heavy jets ejected from the AGN.


\subsection{Radio Mini-Halos}
\label{mini-halo.sec}

\begin{figure*}[ht]
\vspace{0cm}
\centerline{
\includegraphics[width=9cm]{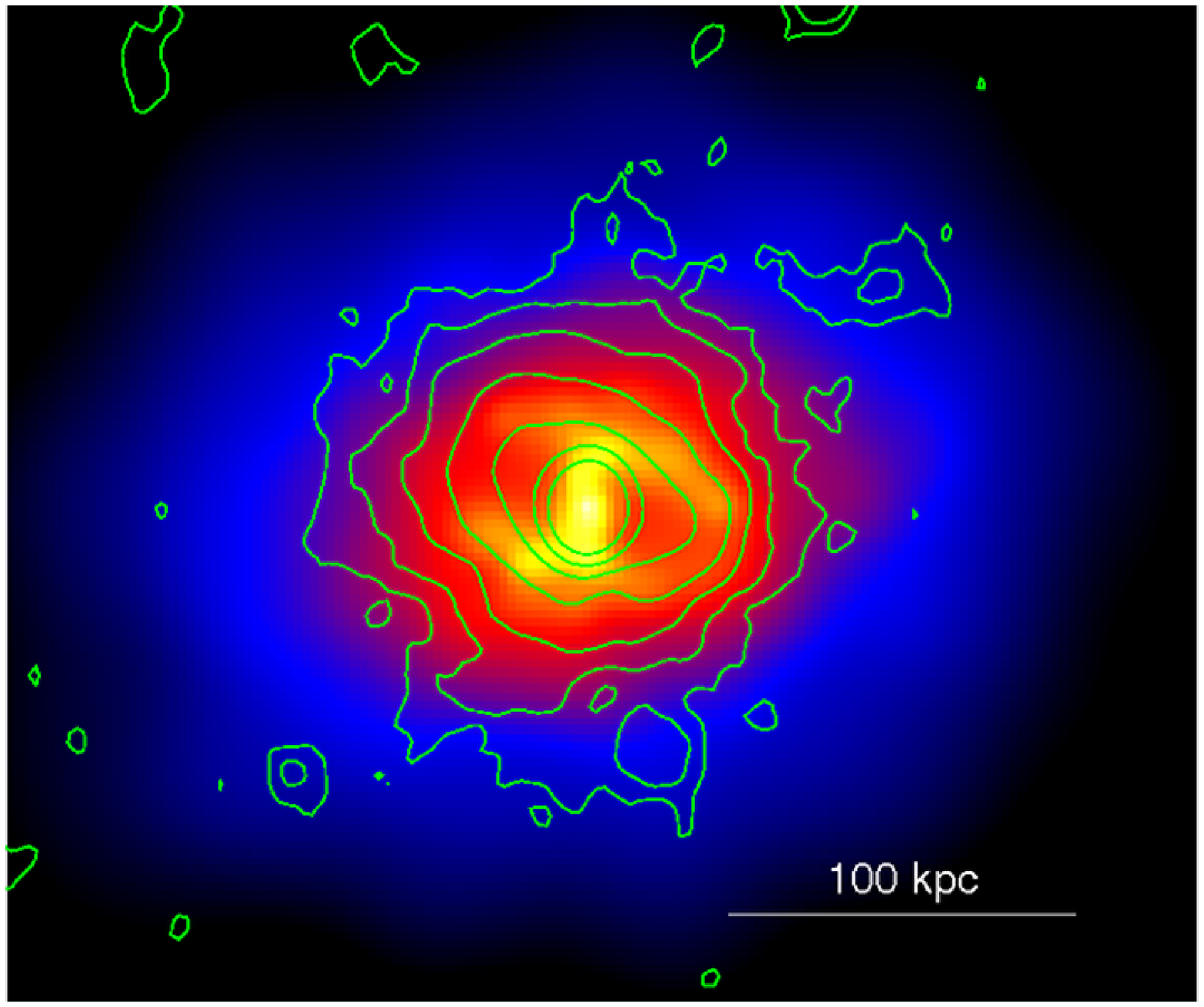}
\includegraphics[width=8cm]{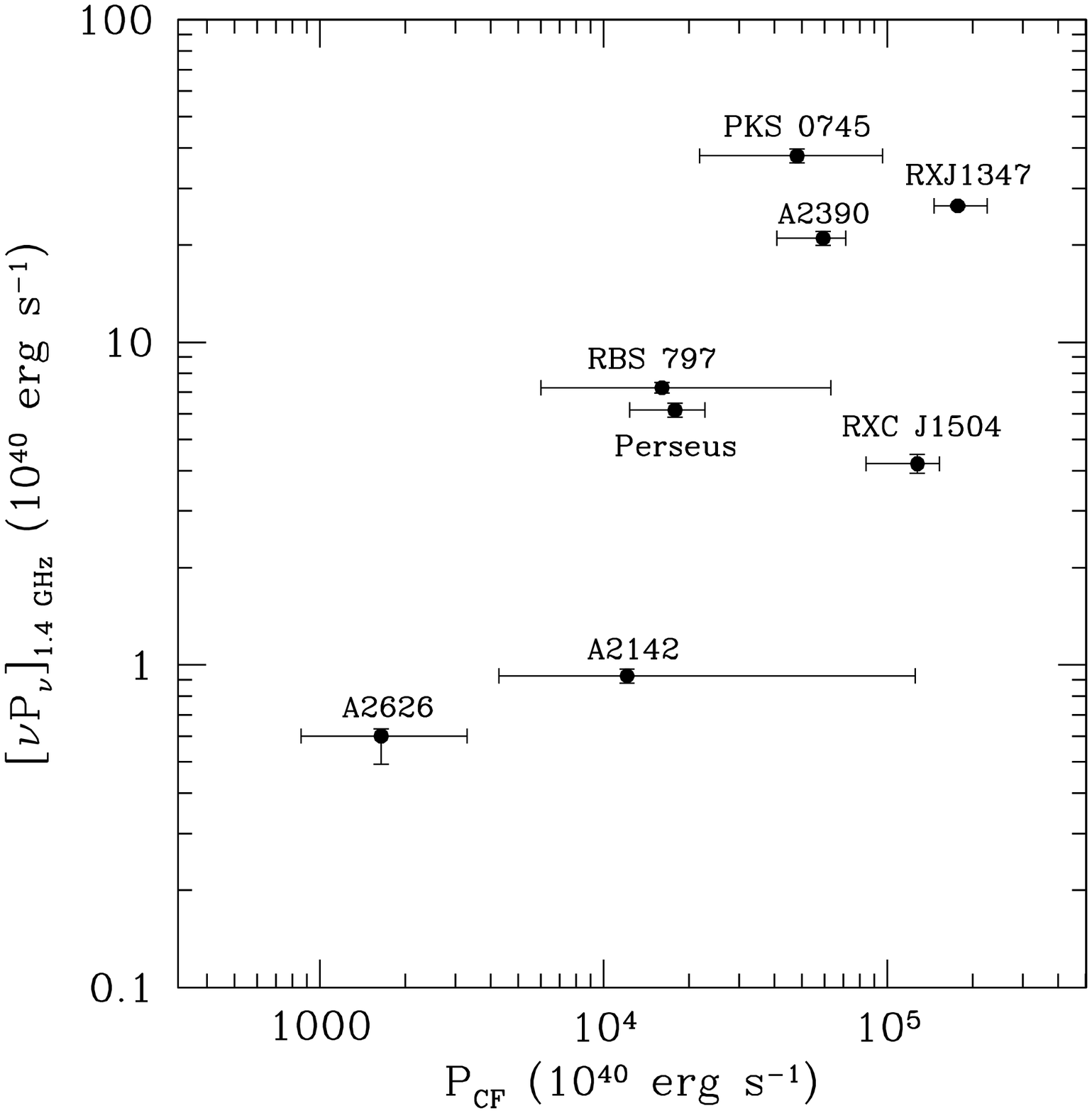}
}
\vspace{0cm}
\caption{\small \label{mini-halo.fig} {\it Left:} 1.4 GHz VLA radio
  contours (obtained by combining observations in A-, B-, and C- array
  configurations) overlaid onto the smoothed 0.5-2.0 keV
  \textit{Chandra} X-ray image of the galaxy cluster RBS 797. The
  combined radio map has a resolution of $3''$, and is able to reveal
  the morphology of the central radio source, showing its elongation
  in the cavity direction (see top right panel of Figure
  \ref{examples.fig}), without losing sensitivity at the larger
  scales. In particular, the extended radio emission is detected out
  to $\sim 90$ kpc. By subtracting the contribution of the central
  nuclear source, the residual flux density of the diffuse radio
  emission is $\simeq 11.5 \pm 0.6$ mJy, indicating the likely
  presence of a radio mini-halo.  {\it Right:} Integrated radio power
  at 1.4 GHz, $\left[ \nu P_{\nu} \right]_{\rm 1.4 GHz}$, vs. cooling
  flow power, $P_{\rm CF} = \dot{M} k T/\mu m_{\rm p}$, for the
  mini-halo clusters which have relevant X-ray and radio data
  available (data from \citep{Gitti_2004, Gitti_2007b, Boeringer_2005,
    Giacintucci_2011a, Doria_2011}).  }
\end{figure*}

In some cases, the powerful radio galaxies at the center of cool core
clusters are surrounded by diffuse radio emission on scales $\sim$
$200-500$ kpc having steep radio spectra ($\alpha > 1 ; S_{\nu}
\propto \nu^{-\alpha}$). These radio sources, generally referred to as
{\it ``radio mini-halos''}, are synchrotron emission from GeV
electrons diffusing through $\mu$G magnetic fields.  Although the
central radio galaxy is the obvious candidate for the injection of the
population of relativistic electrons, mini-halos do appear quite
different from the extended lobes maintained by AGN, therefore their
radio emission proves that magnetic fields permeate the ICM and at the
same time may be indicative of the presence of diffuse relativistic
electrons.  In particular, due to the fact that the radiative lifetime
of radio-emitting electrons ($\sim 10^8$ yr) is much shorter than any
reasonable transport time over the cluster scale, the relativistic
electrons responsible for the extended radio emission from mini-halos
need be continuously re-energized by various mechanisms associated
with turbulence in the ICM (reaccelerated {\it primary} electrons), or
freshly injected on a cluster-wide scale (e.g. as a result of the
decay of charged pions produced in hadronic collisions, {\it
  secondary} electrons).
\citet{Gitti_2002} developed a theoretical model which accounts for
the origin of radio mini--halos as related to electron re-acceleration
by magnetohydrodynamic (MHD) turbulence, which is amplified by
compression in the cool cores.  In this model, the necessary
energetics to power radio mini-halos is supplied by the cooling flow
process itself, through the compressional work done on the ICM and the
frozen-in magnetic field.  The successful application of this model to
two cool core clusters (Perseus: \citep{Gitti_2002} and A 2626:
\citep{Gitti_2004}) has given support to a primary origin of the
relativistic electrons radiating in radio mini-halos.

Radio mini-halos are rare, with only about a dozen objects known so
far.  \citet{Gitti_2004} selected an initial sample of five mini-halo
clusters based on the presence of both a cool core and a diffuse,
amorphous radio emission with no direct association with the central
radio source: Perseus \citep{Burns_1992}, A 2626 \citep{Rizza_2000,
  Gitti_2004}, A 2142 \citep{Giovannini-Feretti_2000}, PKS 0745$-$191
\citep{Baum-O'Dea_1991}, A 2390 \citep{Bacchi_2003}. In these clusters
the size of the diffuse radio emission is comparable to that of the
cooling region. These criteria are now typically adopted to identify
mini-halos.  However, the classification of a radio source as a
mini-halo is not trivial: their detection is complicated by the fact
that the diffuse, low surface brightness emission needs to be
separated from the strong radio emission of the central radio
galaxy. Furthermore, the criteria adopted to define mini-halos are
somehow arbitrary (e.g., total size, morphology, presence of cool
core) and some identifications are still controversial. This said, new
detections of radio mini-halos have recently been claimed in the
galaxy clusters RX J1347.5$-$1145 \citep{Gitti_2007b}, Z 7160
\citep{Venturi_2008}, A 1835 \citep{Govoni_2009}, A 2029
\citep{Govoni_2009}, Ophiuchus \citep{Govoni_2009, Murgia_2010}, RXC
J1504.1$-$0248 \citep{Giacintucci_2011a}, and RBS 797
(Fig. \ref{mini-halo.fig}, left panel, see also \citep{Doria_2011}).

Radio mini-halos are still poorly understood sources.  Although
secondary electron models have been proposed to explain the presence
of their persistent, diffuse radio emission on large-scale in the ICM
\citep[e.g.,][]{Pfrommer-Ensslin_2004, Keshet-Loeb_2010}, a primary
origin of radio mini-halos is now favored by recent statistical
studies \citep{Cassano_2008} and by the observed trend between the
radio power of mini-halos and the maximum power of cooling flows (see
Figure \ref{mini-halo.fig}, right panel). This indicates a direct
connection between cooling flows and radio mini-halos, i.e. the most
powerful radio mini--halos are associated with the most massive
cooling flows, as expected in the framework of the
\citet{Gitti_2002}'s model.  However, the origin of the turbulence
necessary to trigger the electron reacceleration is still debated.
The signatures of minor dynamical activity have recently been detected
in some mini-halo clusters, thus suggesting that additional or
alternative turbulent energy for the reacceleration may be provided by
minor mergers \citep{Gitti_2007b, Cassano_2008} and related gas
sloshing mechanism in cool core clusters
\citep{Mazzotta-Giacintucci_2008, ZuHone_2011}.  Given the prevalence
of mini-halos in clusters with X-ray cavities, another attractive
possibility is that the turbulent energy is provided by a small
fraction of the energy released by the bubbles rising from the central
AGN (as suggested by \citep{Cassano_2008}).  Needless to say, a larger
mini-halo sample as well as further theoretical investigations are
necessary to reach a better understanding of this class of sources.


\subsection{Weak Shocks and Giant Cavities}
\label{giant.sec}

In addition to the cavity enthalpy, shocks driven by the AGN outburst
may contain a large fraction of the energy released, thus working to
heat the ICM.  Such shocks have been long predicted by numerical
simulations \citep{Heinz_2003, Brighenti-Mathews_2006, Gaspari_2011a}
but are difficult to detect since they are relatively weak (with Mach
numbers $\sim 1-2$) and are seen in projection against the cooler,
brighter gas in cluster cores.  We also note that to establish these
surface brightness discontinuities as shocks one must measure an
increase in temperature in the so-called ``post-shock region'', as the
ICM is heated by the passage of the shock . Usually the existing
images are too shallow to rule out, e.g., the possibility that these
features are cold front edges, due to gas sloshing
\citep[e.g.,][]{Markevitch-Vikhlinin_2007}. Besides a very few
examples of strong shocks (e.g., Centaurus A, with Mach number
$\sim$8, \citep{Kraft_2007, Croston_2009}), only recently elliptical
surface brightness edges, consistent with arising from weak shock
fronts driven by the cavities as they initially expanded, have become
to emerge in deep {\it Chandra} exposures of bright clusters and
groups.  Beautiful examples of cocoon shocks are visible in the Hydra
A cluster \citep{Nulsen_2005b, McNamara_2005} and in the NGC 5813
group \citep{Randall_2011}, see left panels of Figure
\ref{examples.fig}.

\begin{figure*}
\vspace{0cm}
\centerline{
\includegraphics[width=15cm]{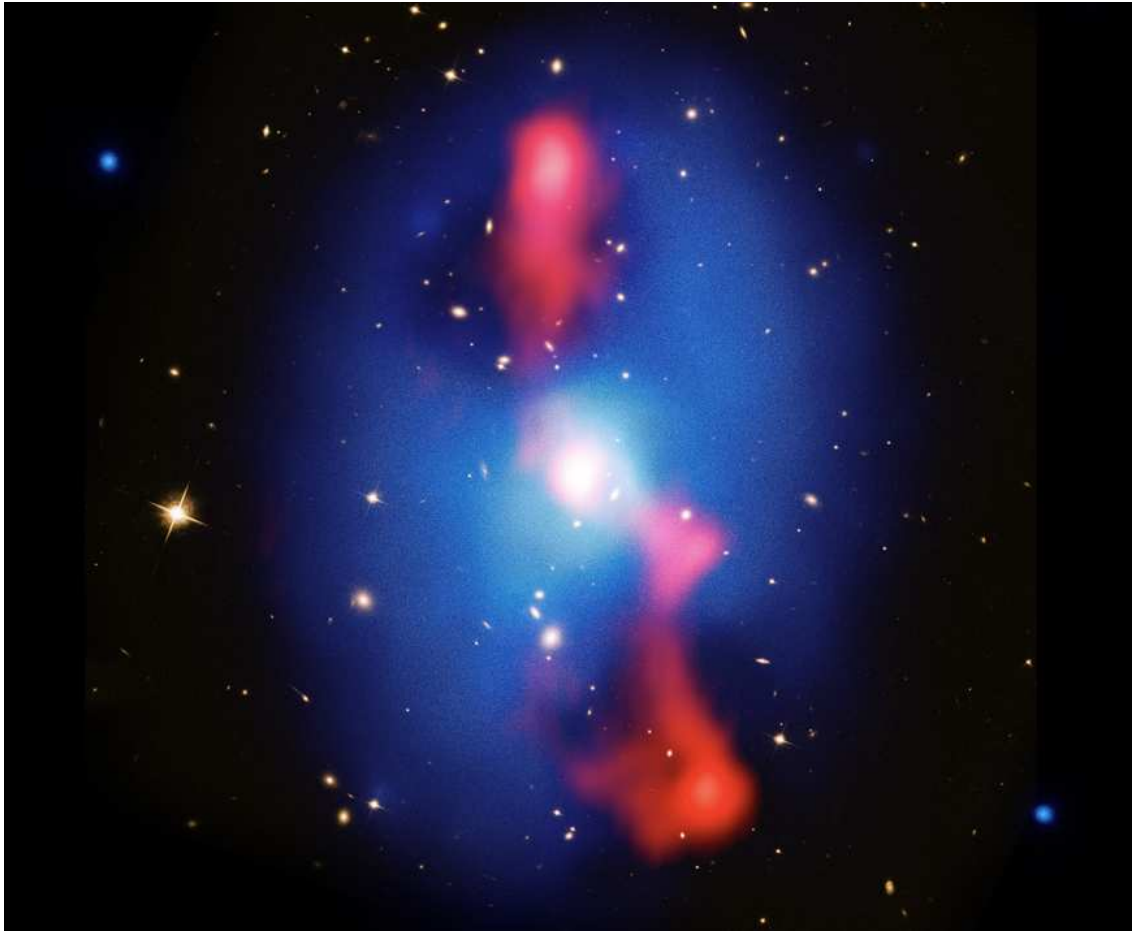}
}
\vspace{0cm}
\caption{\small \label{ms0735.fig} Deep $\sim 500$ ks {\it Chandra}
  X-ray image (blue) and Very Large Array 330 MHz radio image (red)
  superposed with the {\it Hubble Space Telescope} visual image of the
  galaxy cluster MS 0735$+$7421.  The giant X-ray cavities, filled
  with radio emission, are surrounded by a cocoon shock clearly
  visible in the {\it Chandra} image as an elliptical edge. The box is
  roughly 800 kpc by 800 kpc.  }
\end{figure*}

The recent discovery of giant cavities and associated large-scale
shocks in three clusters (MS 0735+7421 \citep{McNamara_2005}, Hercules
A \citep{Nulsen_2005a}, Hydra A \citep{Nulsen_2005b}) has shown that
AGN outbursts not only can affect the central regions, but also have
an impact on cluster-wide scales possibly affecting the global
properties of the ICM and the cluster scaling relations.  In
particular, the giant cavities discovered in the galaxy cluster MS
0735+7421 have a diameter of about 200 kpc each.  The large volume of
the cavities implies a huge cavity power: this large-scale outburst is
the most powerful known so far, releasing upward of $10^{61}$ erg into
the ICM and heating the gas beyond the cooling region
\citep{McNamara_2005, Gitti_2007a}. The new, deep {\it Chandra} image
has confirmed the presence of a weak (Mach number $\sim 1.3$) cocoon
shock surrounding the cavity system (Figure \ref{ms0735.fig}).

\begin{figure*}
\vspace{0cm}
\centerline{
\includegraphics[width=8.3cm]{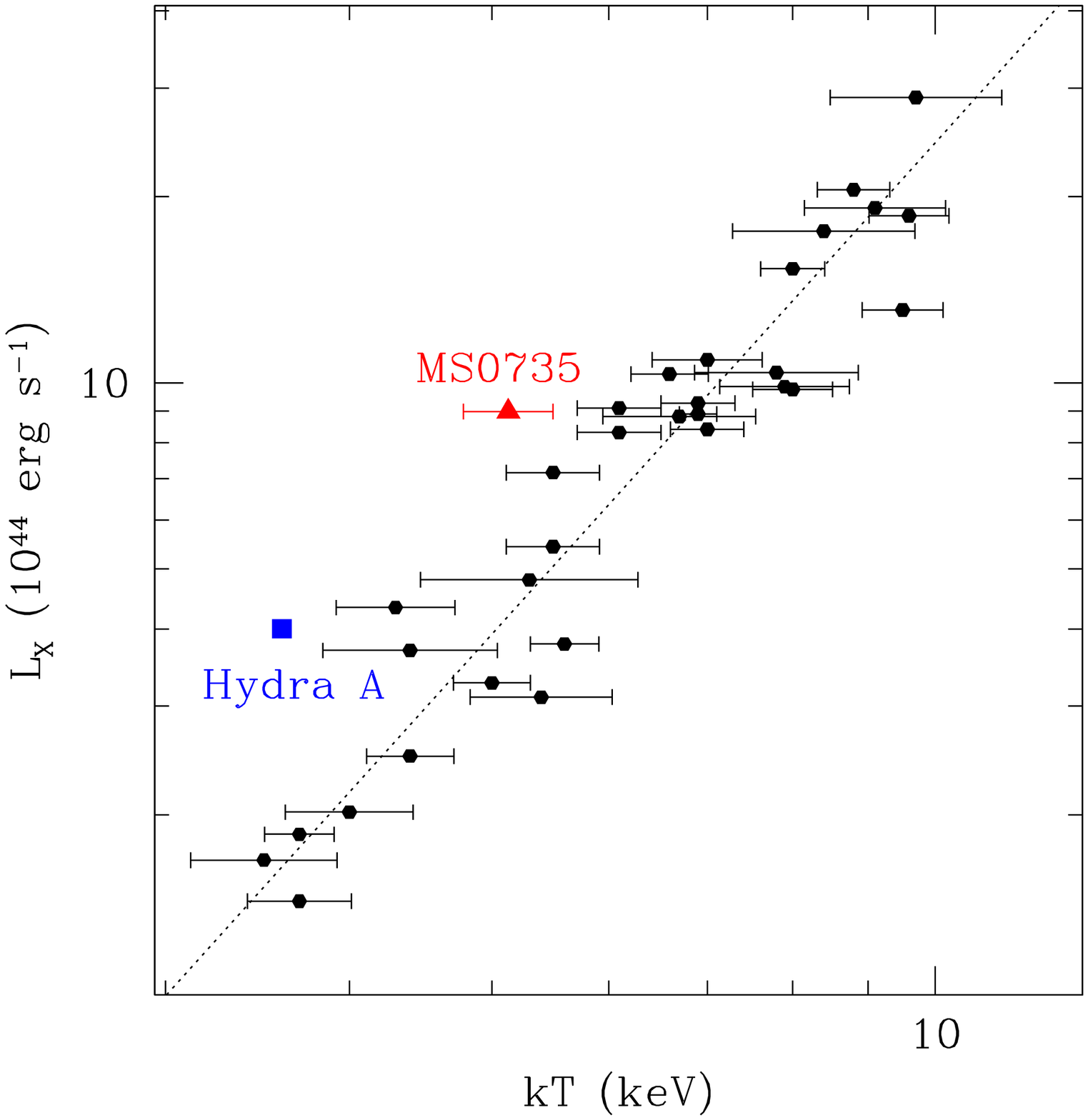}
\includegraphics[width=8.3cm]{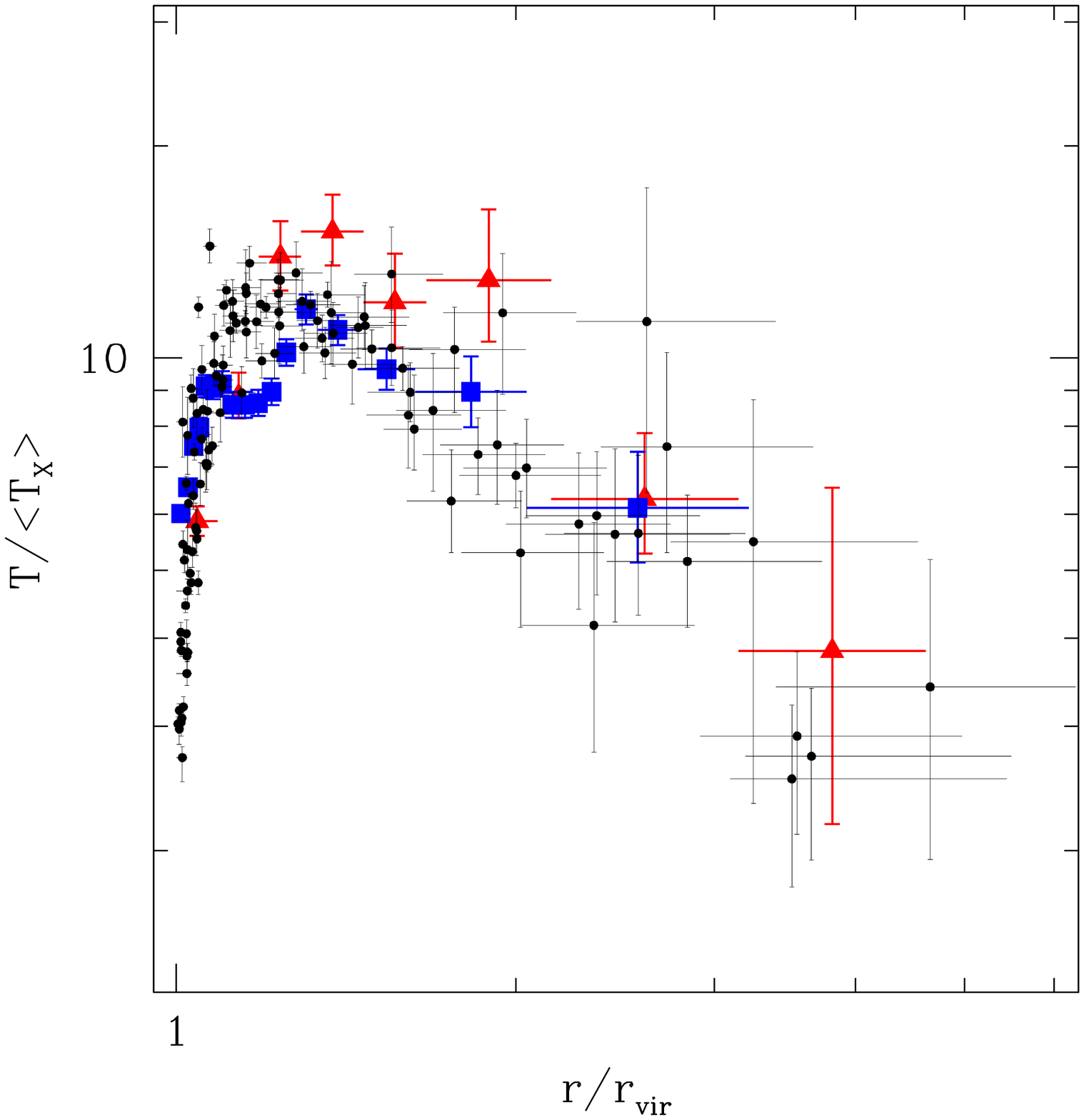}
}
\vspace{0cm}
\caption{\small \label{lt.fig} {\it Left:} Bolometric X-ray
  luminosities corrected for the effect of cooling flow in the central
  $\sim$ 70 kpc vs. emission-weighted temperatures derived excluding
  cooling flow components, from \citet{Markevitch_1998}.  The dashed
  line is the best-fit power law to the sample: $ L_{\rm X} = 6.35
  \cdot (kT/6 \, {\rm keV})^{2.64} \times 10^{44} {\rm erg \, s}^{-1}
  $.  The red triangle and the blue square represent the measurements
  from the {\it XMM-Newton} and {\it Chandra} data of the giant cavity
  clusters MS 0735+7421 and Hydra A, respectively, corrected
  consistently with the method adopted by \citet{Markevitch_1998}.
  {\it Right:} Temperature profiles measured for a sample of relaxed
  clusters presented by \citet{Vikhlinin_2005}.  The temperatures are
  scaled to the cluster emission-weighted temperature excluding the
  central 70 kpc regions.  The profiles for all clusters are projected
  and scaled in radial units of the virial radius $r_{vir}$, estimated
  from the relation $r_{\rm vir} = 2.74 \, {\rm Mpc} \sqrt{<T_{\rm
      X}>/10 \, {\rm keV}}$ \citep{Evrard_1996}.  Overlaid are the
  cooling flow corrected, scaled temperature profiles of the giant
  cavity clusters MS 0735+7421 (red triangles, \citet{Gitti_2007a} and
  Hydra A (blue squares, \citet{Gitti_2011a}).}
\end{figure*}

This new development may have significant consequences for several
fundamental problems in astrophysics.  As seen in \S \ref{cosmo.sec},
the observed relation between X-ray luminosity and gas temperature in
clusters is steeper than expected if cluster growth were governed by
gravity alone.  This steepening is best explained by the addition of
heat to the ICM and is therefore considered the main manifestation of
nongravitational heating.  The discovery of giant cavities has
indicated that powerful AGN outbursts occurring at late times may
contribute a significant fraction of the extra nongravitational
energy.  As mentioned above, this additional heating supplied by AGN
could also induce the suppression of the gas cooling in massive
galaxies required to explain the exponential turnover at the bright
end of the luminosity function of galaxies
\citep[e.g.,][]{Benson_2003}.  This would indicate a common solution
for the two major heating problems associated with the ICM: those of
cooling flow and galaxy formation.  In the case of MS 0735+7421, the
driving energy of the shock as determined using a spherical
hydrodynamic model is $E_{\rm s} \approx 5.7 \times 10^{61}$ erg
\citep{McNamara_2005}.  As estimated by \citet{Gitti_2007a}, the AGN
outburst in this cluster is heating the gas mass within 1 Mpc $(\sim
7.7 \times 10^{13} M_{\odot})$ at the level of about 1/4 keV per
particle, and the heating level increases to $\sim 0.6$ keV per
particle when considering the gas mass within $r_{\rm 2500}$.  This is
a substantial fraction of the 1-3 keV per particle of excess energy
required to heat the cluster \citep{Wu_2000}.  Only a few outbursts of
this magnitude erupting over the life of a cluster would be required
to heat it.
By contrast, MS 0735+7421 is found to be a factor $\sim$2 more
luminous than expected from its average temperature on the basis of
the observed $L$-$T$ relation for galaxy clusters
\citep{Gitti_2007a}. Based on the data presented in
\citet{Gitti_2011a}, we found a similar result for the giant cavity
cluster Hydra A (Figure \ref{lt.fig}, left panel).  Although caution
should be taken in drawing general conclusions from the study of only
a few objects, this indicates that flux limited samples of distant
X-ray clusters may be biased in favor of detecting clusters with
energetic AGN outbursts.  We also note that powerful AGN outbursts may
have a dramatic effect on the gas mass fraction measurements, due to
an overestimate of the gas density \citep{Gitti_2007a}.
The observed departure of MS 0735+7421 and Hydra A from the mean
$L$-$T$ relation is in apparent contradiction with the argument above
that heat should steep the $L-T$ relation, as also indicated by recent
numerical simulations \citep[e.g.,][]{Puchwein_2008}.  However, we
stress that the observed $L-T$ relation is highly dependent on the
definition of the characteristic temperature, i.e. for a fixed
luminosity the position of each point in the plot may vary
significantly depending on the choice of the method adopted to measure
the average emission-weighted temperature for each
cluster. Furthermore, the possibility of building a consistent $L-T$
scaling relation from a sample of clusters relies on the capability to
correct both the temperature and the luminosity measurements for the
effects of the central cooling flow in a consistent manner for the
whole sample.  This may not be trivial as the physical conditions can
vary significantly from case to case. For example, the commonly
adopted method of excluding the central 70 kpc is found to have some
drawback for giant cavity systems as the cooling region and the effect
of AGN feedback extend beyond this radius (see discussion in
\citep{Gitti_2007a}).

On the other hand, based on a study of {\it XMM--Newton} data,
\citet{Gitti_2007a} have shown that the energetic outburst in MS
0735+7421 does not cause dramatic instantaneous departures from the
average properties of the ICM because it has not had a measurable
impact on the large-scale temperature profile, which is in fact
consistent within the scatter of the profiles observed in relaxed
cluster \citep{Vikhlinin_2005}.  We recently found a similar result
(Figure \ref{lt.fig}, right panel) in the Hydra A cluster, although a
sort of {\it ``plateau''} standing below the typical profile in the
range of radius $\sim 0.05-0.1 r_{\rm vir}$ indicates the presence of
cooler gas (see \S \ref{mechanical.sec}).  In general, these results
suggest that there cannot have been many previous outbursts of high
magnitude in these clusters, otherwise the total energy added to the
ICM outside the cooling region should have had a marked effect.
Studies of cavity samples found that the prevalence of outbursts as
energetic as $10^{61}$ erg is 3 (namely, MS 0735+7421, Hercules A and
Hydra A) over 30 \citep{Rafferty_2006}.  If, as it appears from our
in-depth studies of MS 0735+7421 and Hydra A, such powerful outbursts
are rare in individual clusters, their occurrence in $\sim10\%$ of
known cases hence requires that they occur a similar fraction of time
in most cooling flow clusters.  This picture is consistent with the
model proposed by \citet{Nipoti-Binney_2005}, in which the AGN
activity is strongly variable with time and all systems occasionally
experience powerful outbursts.


\subsection{SMBH Growth}
\label{smbh.sec}

AGN are powered by the release of gravitational binding energy from
accretion onto massive black holes \citep{Lynden-Bell_1969,
  Begelman_1984}.  The matter that reaches the black hole converts it
binding energy efficiently into AGN power as $P_{\rm jet} =\epsilon
\dot Mc^2$, where $\epsilon \sim 0.1-0.4$ depending on the spin of the
black hole.  Rapidly spinning black holes with spin parameters
approaching unity are most efficient due to the smaller radius of the
innermost stable circular orbit.  The form of energy that is released
depends on several factors including the accretion rate, the mass of
the black hole, the structure of the accretion flow, and the spin of
the black hole (see \citep{Narayan_2008} for a review).  When black
hole accretion approaches the Eddington rate, the binding energy is
emitted thermally from an optically thick, geometrically thin
accretion disk that is morphologically classified as a quasar or
Seyfert galaxy.  When the accretion rate drops below a few percent of
the Eddington rate a radiatively inefficient AGN is formed (i.e., an
ADAF), and the energy is released primarily in the form of mechanical
energy associated with a radio jet.  Accretion rates in bright AGN can
be estimated using the radiation emitted from the nucleus that
directly (e.g., UV or X-ray emission) or indirectly (eg., nebular or
far-IR emission) trace the energetic output from the accretion disk
\citep[e.g.,][]{Barger_2001}.  However, despite having mechanically
powerful radio AGN \citep{Birzan_2004}, brightest cluster galaxies
(BCGs) rarely show strong X-ray and UV emission emerging from their
nuclei \citep{Hlavacek-Fabian_2011}, implying that their accretion
rates generally lie below a few percent of their Eddington rate.

\citet{Rafferty_2006} estimated the accretion rates in a sample of
BCGs in clusters with prominent X-ray cavities and found this to be
the case.  They estimated the accretion rates using the measured
output power based on the $pV$ work done by the cavities divided by
their buoyancy ages.  Using this approach, \citet{Rafferty_2006} found
that supermassive black holes centred in cool cores are growing at a
rate of $\sim 0.1~M_\odot ~\rm yr^{-1}$.  In rare instances such as
the powerful AGN in MS 0735+7421, \citep{McNamara_2005, McNamara_2009,
  Gitti_2007a}, the accretion rate exceeds $1~M_\odot ~\rm yr^{-1}$.
Assuming black hole masses that are consistent with the values
expected from their stellar luminosities and velocity dispersions, the
accretion rates are consistent with being currently at most a few
percent of Eddington.  If AGN feedback in BCGs quenches cooling flows
over the lifetimes of clusters, their black holes may be more massive
than predicted by the $M_{\rm BH}-\sigma$ relation.


\subsection{Further Evidence for Mechanical Feedback}
\label{mechanical.sec}

\begin{figure*}
\vspace{0cm}
\centerline{
\includegraphics[width=8.0cm]{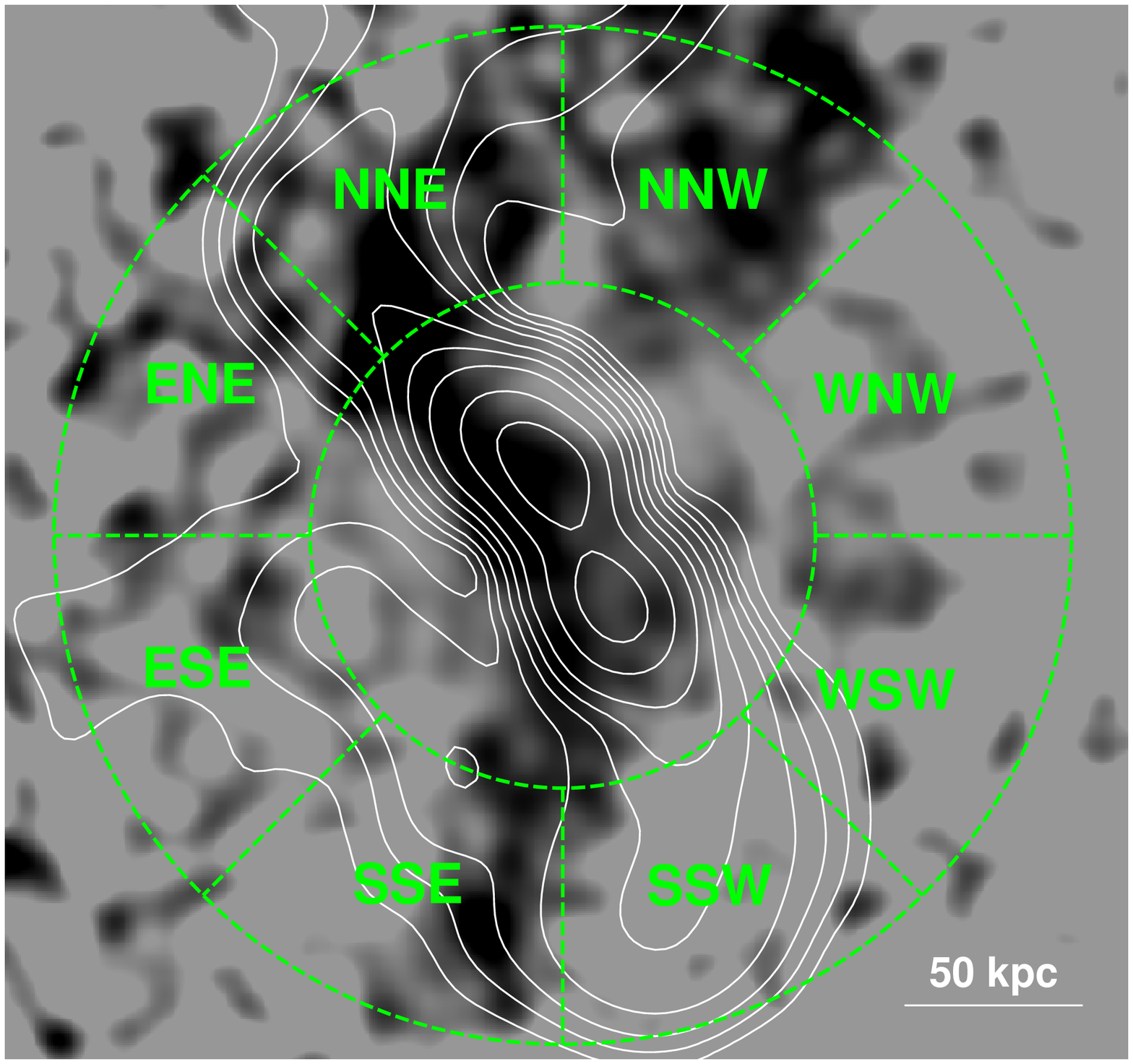}
\includegraphics[width=8.1cm]{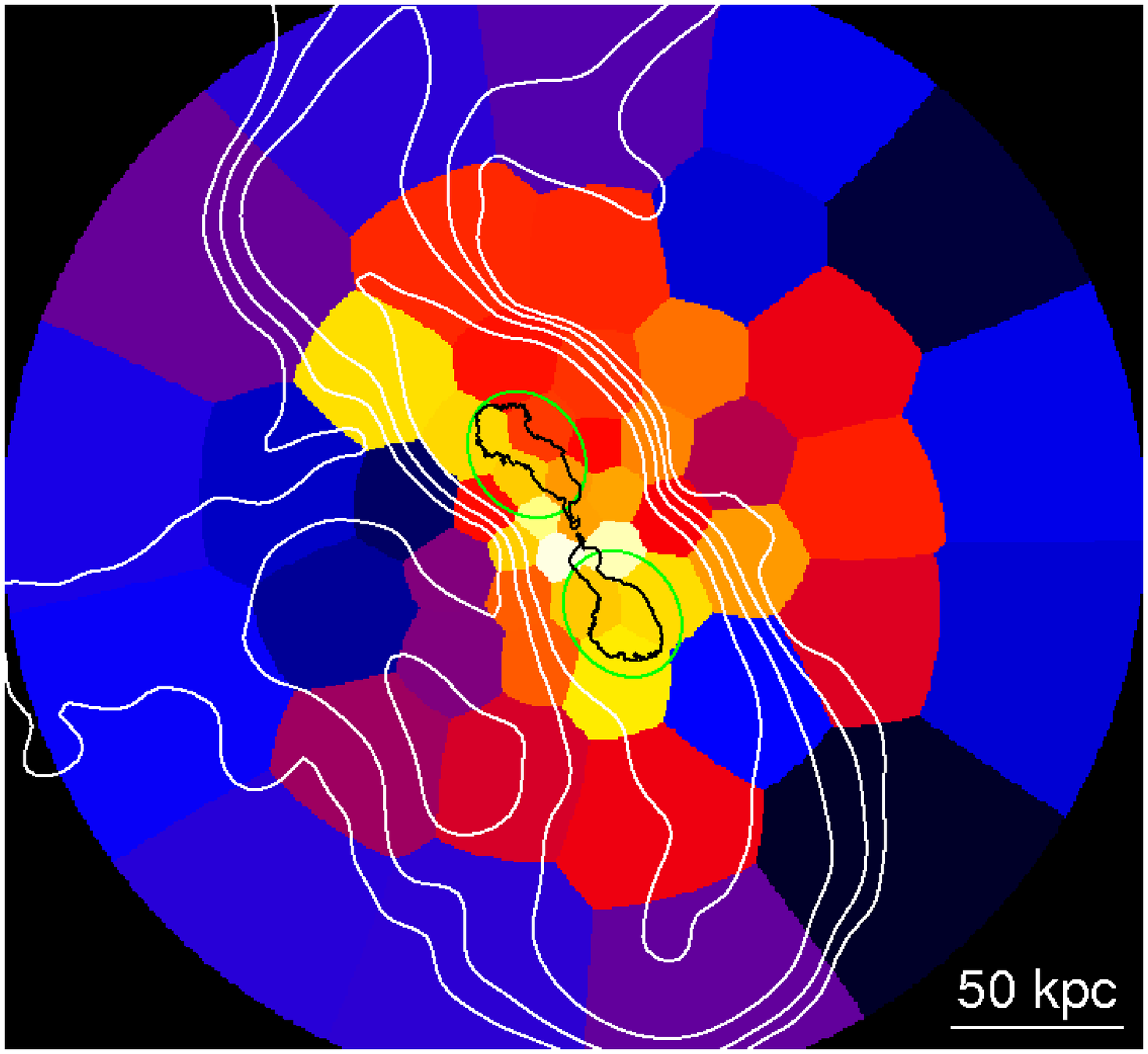}
}
\vspace{0cm}
\caption{\small \label{hydra.fig} {\it Left:} High-contrast hardness
  ratio map of the galaxy cluster Hydra A obtained by dividing a
  1.5-7.5 keV image by a 0.3-1.5 keV image.  Regions in black are
  indicative of low temperature gas, indicating the presence of
  low-entropy filaments. Overlaid in green are the sectors used to
  study the spectral properties of the cool gas (located between radii
  $\sim$ 70-150 kpc.  We extracted the spectra in these sectors and
  compared two different spectral models: a single-temperature plasma
  ``1T'' model, and a ``2T'' model which includes a second thermal
  emission component.  The F statistics for the spectral fitting
  improvement over the single-phase model indicate that the addition
  of a second thermal component is most significant in sectors SSE,
  NNW, NNE, and ENE, thus providing evidence for the presence of
  multiphase gas in agreement with the hardness ratio map.  Adapted
  from \citet{Gitti_2011a}. {\it Right:} Metallicity map showing the
  central $5' \times 5'$ of Hydra A. Brighter regions represent a
  higher metallicity. The color scale of the metallicity $Z$ (in solar
  units) is as follows: white: $Z \geq 0.75$, yellow: $Z=0.65-0.75$,
  orange: $Z=0.55-0.65$, red: $Z=0.45-0.55$, blue: $Z=0.3-0.45$,
  black: $Z \leq 0.3$.  The 1.4 GHz radio emission is shown by the
  black contours. The green elliptical regions indicate the position
  of the inner cavities. Adapted from \citet{Kirkpatrick_2009b}.  In
  both panels are overlaid the white contours outlining the 330 MHz
  radio emission from \citet{Lane_2004}.  }
\end{figure*}

As we have seen in \S \ref{pcav.sec}, it is now widely accepted that
the AGN in the cool cores can reheat the ICM.  Although this is
certainly its main impact, AGN feedback is likely to have other
important effects on the ICM.  We have recently investigated this
point by performing an in-depth study of the galaxy cluster Hydra A,
which harbors a well-known, large-scale system of X-ray cavities
embedded in a ``cocoon'' shock surrounding the central, powerful radio
source (\citep{McNamara_2000, Nulsen_2005b}, see Figure
\ref{examples.fig}, top left panel).  By means of a detailed spectral
analysis of the deep ($\sim 240$ ks) {\it Chandra} observations, we
found indication of the presence of multiphase gas along soft
filaments seen in the hardness ratio map (Figure \ref{hydra.fig}, left
panel). Interestingly, such filaments are almost spatially coincident
with the radio lobes of the powerful central radio source.  The cooler
gas has a significant impact on the radial temperature profile of the
cluster, creating a sort of ``plateau'' which departs from the typical
profile (blue squares in Figure \ref{lt.fig}, right panel). In fact,
the scaled temperature profile of Hydra A measured after masking the
filaments is found to agree well with the general shape of the
temperature profiles observed for relaxed clusters, thus providing a
confirmation that these filaments contain cool gas
\citep{Gitti_2011a}.  By performing a spectral deprojection analysis
of an absorbed 2-temperature component model, we found evidence that
$\sim$10$^{11} M_{\odot}$ of low-entropy material has moved upward
from the central 30 kpc to the observed current position of $75-150$
kpc, likely due to some form of entrainment or dredge up by the rising
lobes.  Assuming that the mass of cool gas, which is $\sim$60\% of the
total mass of gas remaining within 30 kpc \citep{David_2001}, was
lifted out of the central cluster region by a continuous outflow or a
series of bursts from the central AGN over the past $200-500$ Myr (as
it appears from the study of the cavity system, \citep{Wise_2007}), it
would amount to outflows of a few hundred $M_{\odot}$ yr$^{-1}$. There
would thus be a development of gas circulation that can significantly
reduce the net inflow of cooling gas, as initially discussed by
\citet{David_2001} and \citet{Nulsen_2002}. Therefore our results show
that the AGN feedback in Hydra A is acting not only by directly
re-heating the gas, but also by removing a substantial amount of
potential fuel for the supermassive black hole (SMBH).  This provides
indications of mechanical AGN feedback acting through collimated,
massive outflows generated by jets or cavity dragging
\citep[e.g.,][]{Pope_2010b}.

The energy required to lift the cool gas gives a lower limit on the
amount of AGN outburst energy deposited in the ICM.  This value can be
estimated by calculating the variation in gravitational potential
energy during the lifting process.  If we assume that the undisturbed
ICM is approximately isothermal with sound speed $c_{\rm s} \approx
1000$ km s$^{-1}$ and is in a hydrostatic configuration with density
profile $\rho (r)$, we can calculate this quantity as
\citep{Reynolds_2008}
\begin{equation}
  \Delta E = \frac{M_{\rm cool} \, c_{\rm s}^2}{\gamma} \ln 
             \left( \frac{\rho_i}{\rho_f} \right)
\end{equation}
where $M_{\rm cool}$ is the lifted mass, $\rho_i$ and $\rho_f$ are the
initial and final densities of the surrounding ICM, and $\gamma$=5/3
is the ratio of specific heat capacities. From the density profile
presented by David et al. (2001) we estimated that the energy required
to lift the cool gas is \gtsim $2.2 \times 10^{60}$ erg.  This value
is comparable to the work required to inflate all of the cavities
against the surrounding pressure \citep{Wise_2007} and is $\sim$25\%
of the total energy of the large-scale shock \citep{Nulsen_2005b}.  We
also note that a good fraction of the energy required to lift the low
entropy gas will be thermalized when the gas falls back inward.  Given
the large energy required, uplift provides a significant channel for
the dissipation of outburst energy in Hydra A.  There is a remarkable
correlation between these low-entropy filaments and the metal-rich
filaments in the iron-abundance maps measured by
\citet{Simionescu_2009a} and \citet{Kirkpatrick_2009b}, shown in
Figure \ref{hydra.fig} (right panel). The emerging picture is that
Hydra A's powerful radio source is able to lift cool, metal-rich gas
from the central region and distribute it throughout the X-ray
atmosphere of the cluster.  A similar effect of extended metal
outflows in the direction of the radio lobes and X-ray cavities is
observed in other systems (e.g., M87 \citep{Simionescu_2008,
  Werner_2010}, A 262: \citep{Kirkpatrick_2011}, RBS 797
\citep{Doria_2011}).  This is consistent with the results of the most
current theoretical modeling of AGN feedback in massive cosmological
systems, which predict the massive subrelativistic bipolar outflows
and buoyant bubbles to produce a metal uplift along the jet axis (see
\S \ref{simulation.sec}).


\subsection{Numerical Simulations}
\label{simulation.sec}

In the last decade the phenomenon of AGN feedback and its impact on
the ICM has been the subject of many theoretical investigations (see
\citep{Reynolds_2002}, \citep{Ruszkowski-Begelman_2002},
\citep{Brighenti-Mathews_2002}, \citep{Brighenti-Mathews_2003},
\citep{Ruszkowski_2004}, \citep{Jones-DeYoung_2005},
\citep{Soker-Pizzolato_2005}, \citep{Heinz_2006},
\citep{Brighenti-Mathews_2006}, \citep{Bruggen_2007},
\citep{Sternberg_2007}, \citep{Bruggen-Scannapieco_2009},
\citep{Morsony_2010}, \citep{Gaspari_2011a}, \citep{Gaspari_2011b}, to
name a few). These works focused either on several aspects of feedback
physics and microphysics or with the global, long term evolution of
the ICM.  However, uncertainties are still large enough that the
observations must guide researchers to select the relevant mechanisms
at work in the feedback process.

The body of different observational investigations set a number of key
constraints on process of AGN feedback.  The results presented in this
paper strongly suggest that AGN feedback manifests itself as massive
subrelativistic bipolar outflows which heat the ICM through weak
shocks and form X-ray cavities, lift large masses of hot gas from the
central region $\ge 100$ kpc and generate abundance asymmetries along
the outflow direction.  Processes such as AGN Compton heating or
thermal conduction are unable to explain the collection of
observations described above.  Thus, although possibly relevant in
some respect \citep[e.g.,][]{Ruszkowski-Oh_2011}, they likely play a
secondary role in local clusters and groups.

Recent 3D hydro simulations of outflow feedback (\citep{Gaspari_2011a,
  Gaspari_2011b}; see, among others, also \citep{Omma_2004,
  Zanni_2005, Brighenti-Mathews_2006, Sijacki-Springel_2006,
  Vernaleo-Reynolds_2006, Sternberg_2007, Cattaneo_2007, Pope_2010b})
have quantitatively verified that collimated outflows reduce the gas
cooling rate below the limits set by recent {\it Chandra} and {\it
  XMM-Newton} observations (\citep{Peterson-Fabian_2006,
  McNamara-Nulsen_2007} and references therein) for a timescale
comparable with the cluster age. At the same time, this feedback
mechanism generates ICM density and temperature profiles which
reasonably agree with those of typical cool core clusters.  This is
not a trivial result \citep{Brighenti-Mathews_2002,
  Brighenti-Mathews_2003}.  The observable effects of the feedback
from bipolar AGN outflows range from creation of X-ray cavities and
large scale shocks to inducing entropy and metal abundance
anisotropies, due to the lifting of central gas, relatively cool and
metal rich, along the direction of the jet (see also Bruggen 2002;
Roediger et al. 2007). These models, far from being complete and
exhaustive (the origin of the outflows is essentially ad hoc),
represent an interesting starting point for a more thorough
understanding of the AGN feedback process.


\section{~Concluding Remarks}
\label{conclusion.sec}

In order to fully understand the growth and evolution of galaxies and
their central black holes, the history of star formation, and the
formation of large-scale structures, it is crucial to understand first
the processes of cooling, heating and the dynamical evolution of the
intra-cluster gas.  In particular, the feedback from the central black
holes has turned out to be an essential ingredient that must be taken
into account in any model of galaxy evolution.  The main manifestation
of the action of AGN feedback is in galaxy clusters and groups. Their
study, which is currently a very active line of research in
extragalactic astrophysics, has allowed us to make significant
progresses in this field.  However, many details of the AGN feedback
mechanism are still unclear. It is not well understood, for instance,
how the heating distributes in space and time in order to drastically
reduce gas cooling, preserving at the same time the central cool
core. An even more puzzling issue is the process of black hole
accretion and feedback energy generation.

The last decade has represented a quantum leap in the quality of X-ray
observations, thanks to the {\it Chandra} and {\it XMM-Newton}
satellite telescopes. Recent results, discussed in this paper, have
shaked up our understanding of the gas astrophysics in systems ranging
from massive elliptical galaxies to rich galaxy clusters. They suggest
that bipolar outflows emerging from the BCG core inflate large
bubbles, heat the ICM and induce a circulation of gas and metals on
scales of several 100s kpc.

The current generation of X-ray observatories is still working well
and can be expected to continue doing so for few more
years. Unfortunately, the prospects for the future of X-ray astronomy
are not clear at the moment, and X-ray astronomers must rely on the
good health on the existing, but aging, X-ray telescopes.  As a
result, the next few years represent a narrow time window to exploit
the unique opportunity to observe deeply many additional objects, thus
collecting crucial information on the cluster and group evolution.  At
the same time, current radio instrumentation is steadily improving
both at the MHz and at the mm-wave ends of the spectrum, and is about
to make a significant step forward with the next generation of
observatories, such as Low Frequency Array (LOFAR) and Atacama Large
Millimeter/submillimeter Array (ALMA).  A common effort, from both the
observational and theoretical side, will allow us to widen our
knowledge on this fundamental problem which is central to the entire
field of extragalactic astrophysics.


\acknowledgments We especially thank Ewan O'Sullivan for sharing
unpublished results and for providing useful comments. We thank Clif
Kirkpatrick and Scott Randall for permission to print figures from
their work. MG thanks Paul Nulsen, Larry David, Jan Vrtilek, and Ewan
O'Sullivan for many stimulating discussions on this exciting topic,
and Stefano Ettori for carefully reading the original manuscript.  MG
acknowledges support by grants ASI-INAF I/023/05/0 and I/088/06/0 and
by {\it Chandra} grant GO0-11136X.  BRM acknowledges support from the
Natural Sciences and Engineering Research Council of Canada.

\bibliography{bibliography}

\end{document}